\documentclass[11pt,draftcls,onecolumn, romanappendices]{IEEEtran}

\usepackage{amsmath,amsfonts,amssymb,amsthm}
\usepackage{mathrsfs}
\usepackage{dsfont}
\usepackage{comment,blkarray}
\usepackage{multirow,bigdelim}
\usepackage{mathrsfs}
\usepackage{cite}
\usepackage[ruled,commentsnumbered, vlined]{algorithm2e}
\usepackage{tikz}
\usetikzlibrary{automata, arrows}
\usepackage[latin1]{inputenc}
\usepackage{verbatim}
\usepackage{graphicx}
\usepackage{subfigure}		
\usepackage{booktabs}
\usepackage{caption}
\usepackage{algorithmic}
\usepackage{enumerate}

\usepackage{enumitem}

\includecomment{itwfull}
\excludecomment{removeEX4}
\excludecomment{itw2016}
\excludecomment{journalonly}

\usepackage{url}
\usepackage{subfigure}
\usepackage{multicol}

\usepackage{ifpdf}															%
\ifpdf																		%
	\usepackage{hyperref}
\else																		%
\fi	
\usetikzlibrary{arrows,decorations.pathmorphing,decorations.footprints,fadings,calc,
trees,mindmap,shadows,decorations.text,patterns,positioning,shapes,matrix,fit}

\makeatletter
\newcommand{\rmnum}[1]{\romannumeral #1}
\newcommand{\Rmnum}[1]{\expandafter\@slowromancap\romannumeral #1@}

\newif\if@borderstar
\def\bordermatrix{\@ifnextchar*{%
  \@borderstartrue\@bordermatrix@i}{\@borderstarfalse\@bordermatrix@i*}%
}
\def\@bordermatrix@i*{\@ifnextchar[{\@bordermatrix@ii}{\@bordermatrix@ii[()]}}
\def\@bordermatrix@ii[#1]#2{%
\begingroup
  \m@th\@tempdima8.75\p@\setbox\z@\vbox{%
    \def\cr{\crcr\noalign{\kern 2\p@\global\let\cr\endline }}%
    \ialign {$##$\hfil\kern 2\p@\kern\@tempdima & \thinspace %
    \hfil $##$\hfil && \quad\hfil $##$\hfil\crcr\omit\strut %
    \hfil\crcr\noalign{\kern -\baselineskip}#2\crcr\omit %
    \strut\cr}}%
  \setbox\tw@\vbox{\unvcopy\z@\global\setbox\@ne\lastbox}%
  \setbox\tw@\hbox{\unhbox\@ne\unskip\global\setbox\@ne\lastbox}%
  \setbox\tw@\hbox{%
    $\kern\wd\@ne\kern -\@tempdima\left\@firstoftwo#1%
    \if@borderstar\kern2pt\else\kern -\wd\@ne\fi%
    \global\setbox\@ne\vbox{\box\@ne\if@borderstar\else\kern 2\p@\fi}%
    \vcenter{\if@borderstar\else\kern -\ht\@ne\fi%
    \unvbox\z@\kern -\if@borderstar2\fi\baselineskip}%
    \if@borderstar\kern -2\@tempdima\kern2\p@\else\,\fi\right\@secondoftwo#1 $%
  }\null \;\vbox{\kern\ht\@ne\box\tw@}%
\endgroup
}
\makeatother

\newtheorem{thm}{Theorem}

\newtheorem{cor}[thm]{Corollary}

\newtheorem{example}{Example}
\allowdisplaybreaks

\newcommand{\ein}{\mathrm{In}}
\newcommand{\eout}{\mathrm{Out}}
\newcommand{\tail}{\mathrm{tail}}
\newcommand{\head}{\mathrm{head}}

\newcommand{\mincut}{\mathop{\mathrm{mincut}}}

\newcommand{\mN}{\mathcal{N}}
\newcommand{\mG}{\mathcal{G}}
\newcommand{\mV}{\mathcal{V}}
\newcommand{\mE}{\mathcal{E}}
\newcommand{\mW}{\mathcal{W}}
\newcommand{\mbC}{\mathbf{C}}
\newcommand{\mbO}{\bfzero}
\newcommand{\bfzero}{\mathbf{0}}
\newcommand{\hmbC}{\widehat{\mathbf{C}}}
\newcommand{\hmC}{\widehat{\mathcal{C}}}
\newcommand{\hC}{\widehat{C}}
\newcommand{\htheta}{\widehat{\theta}}
\newcommand{\hvarphi}{\widehat{\varphi}}
\newcommand{\mA}{\mathcal{A}}
\newcommand{\mB}{\mathcal{B}}
\newcommand{\mD}{\mathcal{D}}
\newcommand{\mO}{\mathcal{O}}
\newcommand{\mK}{\mathcal{K}}

\newcommand{\mH}{\mathcal{H}}

\newcommand{\vx}{\textit{\textbf{x}}}
\newcommand{\vf}{\textit{\textbf{u}}}
\newcommand{\vY}{\mathbf{Y}}
\newcommand{\vy}{\textit{\textbf{y}}}
\newcommand{\vM}{\mathbf{M}}
\newcommand{\vm}{\textit{\textbf{m}}}

\newcommand{\vK}{\mathbf{K}}
\newcommand{\vk}{\textit{\textbf{k}}}
\newcommand{\vg}{\vec{g}}
\newcommand{\vh}{\vec{h}}
\newcommand{\vb}{\vec{b}}
\newcommand{\va}{\vec{a}}
\newcommand{\Fq}{\mathbb{F}_q}
\newcommand{\Rank}{{\mathrm{Rank}}}

\newcommand{\hg}{\widehat{g}}

\newcommand{\mat}{L}
\newcommand{\matT}{T}
\allowdisplaybreaks
\tikzstyle{vertex}=[draw,circle,fill=gray!30,minimum size=6pt, inner sep=0pt]

\begin{document}
%
\title{Secure Network Function Computation\\ for Linear Functions, Part~\Rmnum{2}: \\Target-Function Security 
}

\author{Yang~Bai,~Xuan~Guang,~and~Raymond~W.~Yeung
}



\maketitle

\begin{abstract}
In this Part~\Rmnum{2} of a two-part paper, we put forward secure network function computation, where in a directed acyclic network, a sink node is required to compute with zero error~a target function of which the inputs are generated as source messages at multiple source nodes, while a wiretapper, who can access any one but not more than one wiretap set in a given collection of wiretap sets, is not allowed to obtain any information about a security function of the source messages. The secure computing capacity for the above model is defined as the maximum average number of times that the target function can be securely computed with zero error at the sink node for one use of the network under the given collection of wiretap sets and security function.
In Part~\Rmnum{1} of the two-part paper, we have investigated securely computing linear functions with the wiretapper who can eavesdrop any edge subset up to a certain size~$r$, referred to as the security level, where the security function is the identity function, i.e., we need to protect the information on source messages from being leaked to the wiretapper. The notion of this security is called {\it source security}. In the current paper, we consider another interesting model which is the same as the above one except that the security function is identical to the target function, i.e., we need to protect the information on the target function from being leaked to the wiretapper. The notion of this security is called {\it target-function security}.
We first prove a non-trivial upper bound on the secure computing capacity, which is applicable to arbitrary network topologies and arbitrary security levels. In particular, when the security level~$r$ is equal to $0$, the upper bound reduces to the computing capacity without security consideration. We further prove that the upper bound is always not less than the upper bound obtained in Part~\Rmnum{1} for source security, which accords with the relation between the secure computing capacities for target-function security and source security. The obtained upper bound depends on the network topology and security level. We further prove an upper bound and a lower bound on this upper bound, which are both in closed form. On the other hand, from an algebraic point of view, we prove two equivalent conditions for target-function security and source security for the existence of the corresponding linear function-computing secure network codes. With them, for any linear function over a given finite field, we develop a code construction of linear secure network codes for target-function security and thus obtain a lower bound on the secure computing capacity; and also generalize the code construction developed in Part~\Rmnum{1} to construct linear secure network codes for source security. We further show that the codes constructed by the generalized code construction for source security (which are evidently target-function secure) are a special subclass of the codes constructed by the code construction for target-function security, and the ratio of the number of codes that can be constructed by the generalized code construction for source security to the number of codes that can be constructed by the code construction for target-function security tends to $0$ as the field size tends to infinity.

\end{abstract}


\IEEEpeerreviewmaketitle


\section{Introduction}

Following Part~\Rmnum{1} of this two-part paper \cite{PartI}, in the current paper we continue to consider \textit{secure network function computation}, which incorporates information-theoretic security with zero-error network function computation. A general setup of the model with single destination is presented as follows. In a directed acyclic graph $\mG$, \rmnum{1}) the single sink node $\rho$ is required to compute with zero error a \textit{target function}~$f$ multiple times, whose arguments are source messages generated at a set of source nodes $S$; \rmnum{2}) a wiretapper, who can access any one but not more than one edge subset $W\in \mW$, is not allowed to obtain any information about a \textit {security function} $\zeta$ of the source messages. Here, $W$ and $\mW$ are referred to as the \textit{wiretap set} and the collection of wiretap sets, respectively. The graph $\mG$, together with $S$ and~$\rho$, forms a \emph{network} $\mN$, and we use the quadruple $(\mN,f,\mW,\zeta)$ to denote the model of secure network function computation. We note that when both the target function $f$ and the security function $\zeta$ are the identity function, the secure model $(\mN,f,\mW,\zeta)$ degenerates to the multi-source single-sink secure network coding model \cite{secure-conference,Cai-Yeung-SNC-IT,Rouayheb-IT,Fragouli-Soljanin-DCC16,GY-SNC-Reduction,Silva-UniversalSNC,GYF-LEP-SNC}; and when no security constraint is considered, the secure model $(\mN,f,\mW,\zeta)$ degenerates to the network function computation model \cite{Koetter-CISS2004,Guang_NFC_TIT19,Giridhar05,Appuswamy11,Appuswamy13,Appuswamy14,YangGuang-TIT18,
Ramamoorthy-Langberg-JSAC13-sum-networks, Rai-Dey-TIT-2012,Kowshik12,Tripathy-Ramamoorthy-IT18-sum-networks,Guang_Zero-Fun-Compre}.

From the information theoretic point of view, we are interested in characterizing the secure computing capacity for $(\mN,f,\mW, \zeta)$, which is defined as the maximum average number of times that the function~$f$ can be securely computed with zero error at the sink node for one use of the network $\mN$ under the consideration of a given collection of wiretap sets $\mW$ and the security function $\zeta$. However, characterizing this secure computing capacity for an arbitrary secure model $(\mN,f,\mW, \zeta)$ is overwhelmingly difficult and complicated, and even for the simpler case of network function computation without any security consideration, the capacity characterization is still open~\cite{Guang_NFC_TIT19}. Thus, in the two-part paper we focus on the model $(\mN,f,\mW, \zeta)$ in which the target function $f$ is a linear function over a finite field and the collection of wiretap sets $\mW$ is the collection of all the edge subsets up to a certain size $r$, referred to as the {\em security level}. We note that linear functions are not only an important class of target functions but also the only class of non-identity functions whose computing capacities over arbitrary network topologies can be characterized when no security constraint is considered.

In Part~\Rmnum{1} of this two-part paper \cite{PartI}, we investigated the case that the security function $\zeta$ is the identity function, namely that we need to protect any information of the source messages from being leaked to the wiretapper. The notion of this security is called \textit{source security}, which is considered in almost all previously studied information-theoretic secure models, e.g., the Shannon cipher system~\cite{Shannon-secrecy}, the secret sharing~\cite{Blakley_secret-sharing-1979,Shamir_secret-sharing-1979}, the wiretap channel~\Rmnum{2}~\cite{wiretap-channel-II}, and secure network coding~\cite{Cai-Yeung-SNC-IT,Rouayheb-IT}. In the current paper, we consider another interesting case that the security function is the same as the target function, i.e., $\zeta=f$. The notion of this security is called \textit{target-function security}. In this case, we need to protect any information on the target function from being leaked to the wiretapper. The main contributions and organization of the paper are given as follows.
\begin{itemize}
  \item In Section~\ref{model_SNFC}, we formally present the model with the target-function-security constraint and define the (function-computing) secure network codes for this model.
  \item In Section~\ref{sec:UppBound}, we first prove a non-trivial upper bound on the secure computing capacity, which is applicable to arbitrary network topologies and arbitrary security levels. In particular, when no security is considered, i.e., $r=0$, the obtained upper bound reduces to the capacity for computing a linear function over a network without security consideration, which we call the \textit{computing capacity} for simplicity. Since the obtained upper bound depends on the network topology and the security level and is not in closed form, we prove an upper bound and a lower bound on this upper bound which are in closed form. Furthermore, we prove that our upper bound on the secure computing capacity is always not less than the upper bound obtained in Part~\Rmnum{1} \cite{PartI} for source security, which accords with the fact that the secure computing capacity for target-function security is always not less than the secure computing capacity for source security.
  \item Section~\ref{subsec_LNC} is devoted to linear (function-computing) secure network coding for the secure model studied in Sections~\ref{model_SNFC} and \ref{sec:UppBound}. We prove an equivalent condition for target-function security for the linear secure network codes from an algebraic point of view. With this equivalent condition, we give in Section~\ref{sec_code_construction} a construction of linear secure network codes for the secure model, which thus implies a lower bound on the secure computing capacity. By the code construction, for any linear function over a given finite field, we can always construct a vector-linear secure network code over the same field for any rate not larger than the obtained lower bound. Finally, an example is given to illustrate our code construction.
  \item Section~\ref{sec:comparison} is devoted to the comparison of our code construction and the code construction for source security discussed in Part~\Rmnum{1}. We first prove an equivalent condition for source security from an algebraic point of view, by which we generalize the code construction developed in Part~\Rmnum{1} \cite{PartI}. By comparing the generalized code construction for source security and our code construction for target-function security, we show that codes constructed by the generalized code construction for source security (which are evidently target-function secure) are a very special subclass of the codes constructed by the code construction for target-function security, and the ratio of the number of codes that can be constructed by the generalized code construction for source security to the number of codes that can be constructed by the code construction for target-function security tends to $0$ as the field size tends to infinity.
  \item In Section~\ref{sec:concl}, we conclude with a summary of our results.
\end{itemize}

\section{Preliminaries}\label{model_SNFC}

\subsection{A Model for Secure Network Function Computation}

We consider a directed acyclic graph $\mG=(\mV, \mE)$, where $\mV$ and $\mE$ are finite sets of nodes and edges, respectively. In $\mG$, multiple edges between two nodes are allowed. We assume that a symbol taken from a finite alphabet $\mB$ can be correctly transmitted on each edge for each use, i.e., we take the capacity of each edge to be $1$ with respect to $\mB$. In the graph $\mG$, we use $\tail(e)$ and $\head(e)$ to denote the \emph{tail} node and the \emph{head} node of an edge $e$, respectively.
For a node $v \in \mV$, we use $\ein(v)$ and $\eout(v)$ to denote the set of input edges and the set of output edges, respectively, i.e., $\ein(v) = \{e \in \mE: \head(e) = v\}$ and $\eout(v) = \{e \in \mE: \tail(e) = v\}$. A sequence of edges $(e_1, e_2, \cdots, e_m)$ is called a (directed) \emph{path} from the edge $e_1$ (or the node $\tail(e_1)$) to the edge $e_m$ (or the node $\head(e_m)$) if $\tail(e_i) = \head(e_{i-1})$ for $i=2, 3, \cdots, m$. In particular, a single edge $e$ is regarded as a path from $e$ to itself (or from $\tail(e)$ to $\head(e)$). For two disjoint subsets of nodes $U$ and $V$, an edge subset $C \subseteq \mE$ is called a \emph{cut} separating~$V$ from $U$ if for each pair $(u, v)$ of $u \in U$ and $v \in V$, no path exists from $u$ to $v$ upon removing the edges in $C$. In particular, for two disjoint singleton subsets of nodes $\{u\}$ and $\{v\}$, a cut separating $\{v\}$ from $\{u\}$ is called a \emph{cut separating $v$ from $u$}. The \emph{capacity} of a cut separating $V$ from $U$ is defined by the size of this cut. A cut $C$ separating $V$ from $U$ is called a \emph{minimum cut} separating $V$ from $U$ if there does not exist a cut $C'$ separating $V$ from $U$ such that $|C'| < |C|$. The capacity of a minimum cut separating $V$ from $U$ on the graph $\mG$ is called the \emph{minimum cut capacity} separating $V$ from $U$, denoted by $\mincut_\mG(U, V)$, or simply $\mincut(U, V)$ when there is no ambiguity on the underlying graph. For two distinct nodes $u$ and $v$ on the graph $\mG$, we similarly define a minimum cut separating $v$ from $u$ and the minimum cut capacity separating $v$ from $u$, written as $\mincut_\mG(u, v)$, or $\mincut(u, v)$ when there is no ambiguity.

In the graph $\mG$, we let $S \subset \mV$ be the set of $s$ \emph{source nodes} $\sigma_1, \sigma_2, \cdots, \sigma_s$ and $\rho \in \mV \setminus S$ be the single \emph{sink node}. Without loss of generality, we assume that each source node $\sigma_i$ has no input edges and the single sink node $\rho$ has no output edges, i.e., $\ein(\sigma_i) = \eout(\rho) = \emptyset, ~i=1, 2, \cdots, s$; and there exists a path from each node $v \in \mV \setminus \{\rho\}$ to $\rho$. The graph $\mG$, together with $S$ and $\rho$, forms a \emph{network} $\mN$, denoted by $\mN = (\mG, S, \rho)$.

Let $f:~\prod_{i=1}^s\mA_i\to \mO$ be a nonconstant function, called the \emph{target function}, that is needed to be computed repeatedly with zero error at the sink node $\rho$, where $\mA_i$, $1\leq i \leq s$ and $\mO$ are all finite alphabets. We assume that the $i$th argument of the target function $f$ is generated at the $i$th source node~$\sigma_i$ for $i=1,2,\cdots,s$. Let $\ell$ and $n$ be two nonnegative integers. We consider computing $f$, the target function, $\ell$ times by using the network $n$ times, i.e., by transmitting at most $n$ symbols in $\mB$ on each edge in $\mE$. For each $i=1,2,\cdots,s$, we let the {\em information source} at the $i$th source node $\sigma_i$ be~a random variable $M_i$ according to the uniform distribution on $\mA_i$. All the sources $M_i$, $1\leq i \leq s$ are mutually independent. Let $M_S=(M_{1},M_{2}, \cdots, M_s)$. The $i$th source node $\sigma_i$ generates $\ell$ independent identical distributed (i.i.d.) random variables $M_{i,1}, M_{i,2}, \cdots, M_{i,\ell}$ with generic random variable $M_i$. We let $\vM_i=(M_{i,1},M_{i,2}, \cdots, M_{i,\ell})$, called the \emph{source message generated by~$\sigma_i$}. We further let $\vM_S=(\vM_1,\vM_2,\cdots,\vM_s)$ be the \emph{source message vector generated by $S$}. The $\ell$ values of the target function $f$
\begin{align}\label{function_f}
f(\vM_S)\triangleq\big(f(M_{1,j},M_{2,j}, \cdots, M_{s,j}): j=1,2,\cdots,\ell \big)
\end{align}
are required to be computed with zero error at $\rho$.

In addition, we consider a collection of edge subsets $\mW$ where each edge subset $W \in \mW$ is called a \emph{wiretap set}, and another nonconstant function $\zeta: \prod_{i=1}^{s}\mA_i \rightarrow \mathcal{Q}$, called the {\em security function}, where $\mathcal{Q}$ is the image set of $\zeta$. In this model, the $\ell$ values of the target function $f(\vM_S)$ are required to be correctly computed at $\rho$ through the network $\mN$, while the $\ell$ values of the security function $\zeta$
\begin{align*}
  \zeta(\vM_S) \triangleq \big( \zeta(M_{1, j}, M_{2, j}, \cdots, M_{s, j}): ~j=1, 2, \cdots, \ell \big)
\end{align*}
are required to protect from being leaked to a wiretapper who can access any one but not more than one wiretap set in $\mW$. The collection of wiretap sets $\mW$ and the security function $\zeta$ are known by the source nodes and the sink node but which wiretap set in $\mW$ is eavesdropped by the wiretapper is unknown. We use $(\mN,f,\mW,\zeta)$ to denote the model as specified above, called the model of secure network function computation.


\subsection{Function-Computing Secure Network Code for  $(\mN, f, \mW, \zeta)$}

In this subsection, we will define (function-computing) secure network codes for the model of secure network function computation $(\mN, f, \mW, \zeta)$. The definition of the codes is similar to the definition of secure network codes for the model discussed in Part~\Rmnum{1} of the current work~\cite{PartI}.

As a part of the secure network code to be defined, we assume that for $i=1, 2, \cdots, s$, a random variable~$\vK_i$, called the \emph{key}, which is uniformly distributed on a finite set $\mK_i$,\footnote{If no randomness is needed at the source node $\sigma_i$, then we let $\mK_i = \emptyset$. Here, the unnecessariness of randomness is possible for the function security.} is available to the source node~$\sigma_i$. We let $\vK_S = (\vK_1, \vK_2, \cdots, \vK_s)$. Further, assume that all the keys $\vK_i$ and the source messages $\vM_i, i=1, 2, \cdots, s$ are mutually independent. We now consider securely computing $f$, the target function,~$\ell$ times by using the network $n$ times under the collection of wiretap sets $\mW$ and the security function~$\zeta$. An $(\ell, n)$ \emph{(function-computing) secure network code} for the model $(\mN, f, \mW, \zeta)$ is defined as follows. First, we let $\vm_i \in \mA_i^\ell$ and $\vk_i \in \mK_i$ be arbitrary outputs of the source message $\vM_i$ and the key~$\vK_i$, respectively, for $i=1, 2, \cdots, s$. Accordingly, let $\vm_S = (\vm_1, \vm_2, \cdots, \vm_s)$ and $\vk_S = (\vk_1, \vk_2, \cdots, \vk_s)$, which can be regarded as two arbitrary outputs of $\vM_S$ and $\vK_S$, respectively. An $(\ell, n)$ secure network code $\hmbC$ consists~of
\begin{itemize}
  \item a \emph{local encoding function} $\htheta_e$ for each edge $e \in \mE$ such that
\begin{align}\label{def_local_encoding}
  \htheta_e:
  \begin{cases}
    \qquad\mA_i^\ell \times \mK_i \rightarrow \mB^n & \mbox{if } \tail(e) = \sigma_i \mbox{ for some } i,\\
    \prod\limits_{d \in \ein(\tail(e))}\mB^n \rightarrow \mB^n & \mbox{otherwise},
  \end{cases}
\end{align}
where the local encoding functions $\htheta_e,~\forall~e \in \mE$ are used to compute the messages transmitted on the edges and are executed following a given topological order on the edges in $\mE$;
  \item a \emph{decoding function} $\hvarphi: \prod_{\ein(\rho)}\mB^n \rightarrow \mO^\ell$ at the sink node $\rho$, which is used to compute the target function $f$ with zero error.
\end{itemize}

We let $\vy_e \in \mB^n$ be the message transmitted on each edge $e \in \mE$ by using the code $\hmbC$ with the source message vector $\vm_S$ and the key vector $\vk_S$. With the encoding mechanism as described in \eqref{def_local_encoding}, we readily see that $\vy_e$ is a function of $\vm_S$ and $\vk_S$, denoted by $\hg_e\big(\vm_S, \vk_S\big)$ (i.e., $\vy_e = \hg_e\big(\vm_S, \vk_S\big)$), where $\hg_e$ can be obtained by recursively applying the local encoding functions $\htheta_e, e \in \mE$. More precisely, for each $e \in \mE$,
\begin{align*}
  \hg_e\big(\vm_S~ \vk_S\big) =
  \begin{cases}
    \htheta_e\big(\vm_i~ \vk_i\big) & \mbox{if } \tail(e) = \sigma_i \mbox{ for some } i,\\
    \htheta_e\big(\hg_{\ein(u)}(\vm_S~ \vk_S)\big) & \mbox{otherwise},
  \end{cases}
\end{align*}
where $u=\tail(e)$ and $\hg_E(\vm_S~ \vk_S) = \big(\hg_e(\vm_S~ \vk_S): e \in E\big)$ for an edge subset $E \subseteq \mE$. We call $\hg_e$ the \emph{global encoding function} of the edge $e$ for the code $\hmbC$.

For the model $(\mN, f, \mW, \zeta)$, we say an $(\ell, n)$ secure network code $\hmbC = \{\htheta_e: e \in \mE;~\hvarphi\}$ is \emph{admissible} if the following \emph{computability} and \emph{function-security conditions} are satisfied:
\begin{itemize}
  \item \textbf{\emph{computability condition}}: the sink node $\rho$ computes the target function $f$ with zero error, i.e.,
      \begin{align*}
        \hvarphi \big(\hg_{\ein(\rho)}(\vm_S~ \vk_S)\big) = f(\vm_S),\quad \forall~\vm_S \in \prod_{i=1}^{s}\mA_i^\ell \text{~~and~~} \vk_S \in \prod_{i=1}^{s}\mK_i;
      \end{align*}
  \item \textbf{\emph{function-security condition}}: for each wiretap set $W \in \mW$, $\vY_W$ and $\zeta(\vM_S)$ are independent, i.e.,
      \begin{align}\label{def_sec_condition}
        I\big(\vY_W;\zeta(\vM_S)\big) = 0,
      \end{align}
      where $\vY_W = (\vY_e: e \in W)$ with $\vY_e \triangleq \hg_e(\vM_S, \vK_S)$ being the random vector transmitted on the edge~$e$.
\end{itemize}
The \emph{secure computing rate} of such an admissible $(\ell, n)$ secure network code $\hmbC$ is defined by
\begin{align*}
  R(\hmbC) = \frac{\ell}{n},
\end{align*}
which describes the average number of times the function $f$ can be securely computed with zero error at $\rho$ for one use of the network $\mN$, while no information about the security function $\zeta$ is leaked to the wiretapper who can access any one but not more than one wiretap set $W \in \mW$. Further, we say a nonnegative real number~$R$ is called \emph{achievable} if for all $\epsilon > 0$, there exists an admissible $(\ell, n)$ secure network code $\hmbC$ for the model $(\mN, f, \mW, \zeta)$ such that
\begin{align*}
  R(\hmbC)  = \frac{\ell}{n} > R - \epsilon.
\end{align*}
Accordingly, the {\em secure computing rate region} for the secure model $(\mN, f, \mW, \zeta)$ is defined as
\begin{align*}
\mathfrak{R}(\mN, f, \mW, \zeta) = \Big\{ R:~R \text{ is achievable for $(\mN, f, \mW, \zeta)$} \Big\},
\end{align*}
and the {\em secure computing capacity} for $(\mN, f, \mW, \zeta)$ is defined as
\begin{align*}
\hmC(\mN, f, \mW, \zeta) =  \max~\mathfrak{R}(\mN, f, \mW, \zeta).
\end{align*}

We end this subsection with an interesting observation that randomness, sometimes, is unnecessary to guarantee the function security in the secure model $(\mN, f, \mW, \zeta)$. This is different from the case of the classical information-theoretic secure models, where randomness is necessary to guarantee the security of the information sources. We take the following example to show this.

\begin{example}\label{example_toy}
  We consider computing the algebraic sum $f(m_1, m_2) = m_1 + m_2$ over the finite field $\mathbb{F}_2$ on the graph $\mG$ as depicted in Fig.~\ref{fig_example_nokey}, while the wiretapper, who can access $e_1$ or $e_2$, attempts to obtain some information on this algebraic sum. A $(1, 1)$ secure network code without randomness is given in Fig.~\ref{fig_example_nokey}, where we can readily see that the computability and security conditions are satisfied.

\begin{figure}[!t]
  \centering
{
 \begin{tikzpicture}[x=0.6cm]
    \draw (-1.8,0) node[vertex] (1) [label=above:$\sigma_1:m_1$] {};
    \draw ( 1.8,0) node[vertex] (2) [label=above:$\sigma_2:m_2$] {};
    \draw ( 0,-2) node[vertex] (3) [label=below: $\rho:m_1 + m_2$] {};

    \draw[->,>=latex] (1) -- (3) node[pos=0.3, auto, right=-0.7mm] {$m_1$};
    \draw[->,>=latex] (1) -- (3) node[pos=0.7, auto, left=0.5mm] {$e_1$};
    \draw[->,>=latex] (2) -- (3) node[pos=0.3, auto, left=-0.5mm,] {$m_2$};
    \draw[->,>=latex] (2) -- (3) node[pos=0.7, auto, right=0.5mm,] {$e_2$};
    \end{tikzpicture}
}
\caption{A toy example to show the unnecessariness of randomness.}\label{fig_example_nokey}
\end{figure}
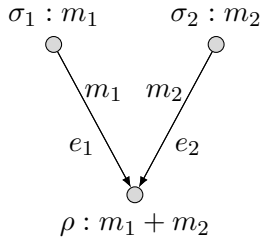
\end{example}

\subsection{Secure Network Function Computation for Linear Functions}

Following Part~\Rmnum{1} of the current work~\cite{PartI}, we continue to consider the case that the target function~$f$ is a \emph{linear function} over a finite field, i.e.,
\begin{align*}
  f(m_1, m_2, \cdots, m_s) = \sum_{i=1}^{s}a_i \cdot m_i, 
\end{align*}
where $m_i, a_i \in \Fq$ for all $i=1, 2, \cdots, s$ (i.e., $\mA_i = \mO = \Fq,~i=1,2 , \cdots, s$), and the collection of wiretap sets $\mW$ is
\begin{equation*}
  \mW_r \triangleq \{W \subseteq \mE: 0 \leq |W| \leq r\},\footnote{We remark that the empty set $\emptyset$, regarded as the wiretap set of size $0$, is in $\mW_r$.}
\end{equation*}
i.e., the wiretapper is able to eavesdrop any one edge subset up to a size $r$, referred to as the \emph{security level}. Further, we let the security function $\zeta$ be the linear function $f$, i.e., $\zeta=f$. With this, the function-security condition \eqref{def_sec_condition} becomes
\begin{equation}\label{eq_sec_condition_for_f}
  I\big(\vY_W;f(\vM_S)\big) = 0,~~\forall~W \in \mW_r,
\end{equation}
which we call the {\it target-function-security} condition for more preciseness. For notational simplicity, we write the secure model $(\mN, f, \mW_r, f)$ as $\langle \mN, f, r \rangle$.\footnote{We mention the difference between the two secure models $\langle \mN, f, r \rangle$ and $(\mN, f, r)$, where in the latter model $(\mN, f, r)$, the security function is the identity function (i.e., source security), namely that all the information sources are required to be protected from being leaked to the wiretapper (cf.~\cite{PartI}).} Accordingly, the (target-function-security) secure computing capacity $\hmC(\mN, f, \mW_r, f)$ is written as $\hmC\langle \mN, f, r \rangle$. Without loss of generality, we further assume  that $\mB = \Fq$, i.e., an element in the field $\Fq$ can be transmitted on each edge for each use.

\section{Upper Bound on the Secure Computing Capacity}\label{sec:UppBound}

Similar to the discussion in~\cite[Section II]{PartI}, the characterization of the capacity for securely computing a linear function over a finite field while protecting this linear function is equivalent to the characterization of the capacity for securely computing an algebraic sum over the same field while protecting this algebraic sum. As such, in the rest of the paper, we let the target function $f$ be the algebraic sum over a finite field~$\Fq$, i.e.,
\begin{align*}
f(m_1,m_2,\cdots,m_s)=\sum_{i=1}^{s} m_i.
\end{align*}

\begin{journalonly}
We consider the secure model $\langle \mN, f, r \rangle$ with $f$ being a linear function over a finite field $\Fq$, i.e.,
\begin{align*}
f(m_1,m_2,\cdots,m_s)=\sum_{i=1}^{s}a_i \cdot m_i.
\end{align*}
For the indices $i$ with $a_i=0$, we remove the terms $a_i \cdot m_i$ from $f$ to form a new linear function $f'$, and at the same time remove the source nodes $\sigma_i$ together with the output edges in $\eout(\sigma_i)$ from the graph~$\mG$ to form a new graph $\mG'$. We update the network $\mN$ to a new one $\mN'=\big( \mG', S\setminus \{ \sigma_i:~a_i=0 \}, \rho \big)$. We readily see that with the security level $r$, securely computing $f$ over $\mN$ and protecting $f$ is equivalent to securely computing $f'$ over $\mN'$ and protecting $f'$. Thus, $\hmC\langle \mN, f, r \rangle=\hmC\langle \mN', f', r \rangle$. {\color{blue}Accordingly, we now assume that in the model $\langle \mN, f, r \rangle$, the linear function $f$ is $f(m_1,m_2,\cdots,m_s)=\sum_{i=1}^{s}a_i \cdot m_i$ with $a_i \neq 0$ for all $1\leq i \leq s$. We let $x_i=a_i \cdot m_i$ and consider the algebraic sum $g(x_1,x_2,\cdots,x_s)=\sum_{i=1}^{s}x_i$ over $\Fq$. Hence, an admissible $(\ell,n)$ secure network code for the model $\langle \mN, f, r \rangle$ can be readily modified into an admissible $(\ell, n)$ secure network code for the model $\langle \mN, g, r \rangle$, and vice versa.} So, we have $\hmC\langle \mN, f, r \rangle=\hmC\langle \mN, g, r \rangle$. We thus have shown that in order to investigate the secure network function computation for a linear function, it suffices to consider secure network function computation for an algebraic sum. Accordingly, in the rest of the paper, we let the target function $f$ be the algebraic sum over a finite field~$\Fq$, i.e.,
\begin{align*}
f(m_1,m_2,\cdots,m_s)=\sum_{i=1}^{s} m_i.
\end{align*}
\end{journalonly}

\subsection{The Upper Bound}

We first present some graph-theoretic notations following the ones in \cite{PartI}. Given a set of edges $C \subseteq \mE$, we define three subsets of source nodes:
\begin{align}
 D_C & =  \big\{ \sigma\in S:~\exists\ e\in C \text{ s.t. } \sigma\rightarrow e \big\},\label{eq_notation_DIJ}\\
 I_C & = \big\{ \sigma\in S:~ \sigma \nrightarrow \rho \text{ upon deleting the edges in $C$ from $\mathcal{E}$} \big\},\nonumber\\
 J_C & = D_C \setminus  I_C,\nonumber
\end{align}
where $\sigma \rightarrow e$ denotes that there exists a path from $\sigma$ to $e$, and $\sigma \nrightarrow \rho$ denotes that there exists no path from $\sigma$ to $\rho$. We can readily see that $I_C \subseteq D_C$. Here,~$J_C$ is the subset of the source nodes $\sigma$ satisfying that there exists not only a path from $\sigma$ to $\rho$ passing through an edge in $C$ but also a path from $\sigma$ to $\rho$ not passing through any edge in~$C$. Further, an edge set~$C$ is said to be a {\em cut set} if and only if $I_C\neq \emptyset$, and we let $\Lambda(\mathcal{N})$ be the family of all the cut sets in the network~$\mathcal{N}$, i.e.,
\begin{align*}
\Lambda(\mathcal{N})=\big\{ C\subseteq \mE:\ I_C \neq \emptyset \big\}.
\end{align*}
In particular, we say a cut set $C$ is a {\em global cut set} if $I_C=S$.

For a cut set $C \in \Lambda(\mN)$ and a wiretap set $W \in \mW_r$, the pair $(C, W)$ is said to be \emph{valid} if either one of the following two conditions is satisfied:

\noindent\textbf{cond 1:} $W \subseteq C$ and $I_C \setminus D_W \neq \emptyset$;

\noindent\textbf{cond 2:} $W \subseteq C$ and $D_W = I_C = S$ (i.e., $I_C \setminus D_W = \emptyset$ and $I_C = S$).

This implies that for such a valid pair $(C, W)$, either there exists $\sigma\in I_C$ such that $\sigma \nrightarrow e$ for all $e\in W$, or $C$ is a global cut set and for each $\sigma\in S=I_C$, there exists $e\in W$ such that $\sigma \rightarrow e$.
We now present our upper bound on the secure computing capacity $\hmC\langle \mN, f, r \rangle$ in the following theorem.

\begin{thm}\label{thm_general_upper}
Consider the secure model $\langle \mN, f, r \rangle$, where the target function~$f$ is an algebraic sum over a finite field $\Fq$. Then,
\begin{align}\label{eq_upper_algebraic_sum}
   \hmC\langle \mN, f, r \rangle \leq
   \min_{ (C, W)\in \Lambda(\mN) \times \mW_r :\atop  (C, W)\,\text{\rm is valid}} \big( |C|-|W| \big).
   \end{align}
\end{thm}
\begin{IEEEproof}
We let $\hmbC$ be any admissible $(\ell,n)$ secure network code for the secure model $\langle \mN, f, r \rangle$, of which the global encoding functions are $\hg_e$, $e\in\mE$. We consider any cut set $C \in \Lambda(\mN)$, i.e., $I_C\neq \emptyset$. Recall that all the source messages $M_{i,j}$ for $i=1,2,\cdots, s$ and $j=1,2,\cdots,\ell$ are i.i.d. according to the uniform distribution on $\Fq$. Thus, we have
\begin{align}\label{equ1}
H\left(\,\sum_{i\in I_C}\vM_i\right)=\ell \cdot H\left(\,\sum_{i\in I_C} M_{i} \right)=\ell \cdot \log q,\footnotemark
\end{align}
where we recall that $\vM_i$ is the vector of the $\ell$ random variables $M_{i,j}$, $1\leq j \leq \ell$ with generic random variable $M_i$, i.e., $\vM_i=(M_{i,1},M_{i,2}, \cdots, M_{i,\ell})$, and similar to \eqref{function_f}, we let
\begin{align*}
\sum_{i\in I_C}\vM_i\triangleq \left(\,\sum_{i\in I_C}M_{i,j} : j=1,2,\cdots,\ell \right).
\end{align*}
\footnotetext{Here, for a subset of the source nodes $I\subseteq S$, we also use $I$ to represent the index set $\{i:~\sigma_i\in I\}$. This abuse of notation should cause no ambiguity.}

We first prove the equality
\begin{align}\label{pf-equ5}
H \left(\,\sum_{i\in I_C}\vM_i \Big| \vY_{C},\vM_{S\setminus I_C},\vK_{S\setminus I_C} \right)=0,
\end{align}
where for a subset $I \subseteq S$, we use $\vM_I$ and $\vK_I$ to denote $(\vM_i:~i \in I)$ and $(\vK_i:~i \in I)$, respectively,~i.e.,
\begin{equation*}
  \vM_I = (\vM_i:~i \in I)~~\text{and}~~\vK_I = (\vK_i:~i \in I).
\end{equation*}
We consider the edge subset
$C' \triangleq \bigcup_{i\in {S\setminus I_C}} \eout(\sigma_i)$. Then,
$D_{C'}=I_{C'}=S\setminus I_C$. So, we see that $\vY_{C'}$ depends only on $\vM_{S\setminus I_C}$ and $\vK_{S\setminus I_C}$, and we can write
\begin{align*}
\vY_{C'}=\hg_{C'}(\vM_S,\vK_S)=\hg\,'_{C'}(\vM_{S\setminus I_C},\vK_{S\setminus I_C}).
\end{align*}
By the above discussion, we have
\begin{align}\label{pf-equ10}
\begin{split}
H\left(\,\sum_{i\in I_C}\vM_i \Big| \vY_{C},\vM_{S\setminus I_C},\vK_{S\setminus I_C}\right)
&=H\left(\,\sum_{i\in I_C}\vM_i \Big| \vY_{C},\vY_{C'},\vM_{S\setminus I_C},\vK_{S\setminus I_C}\right)\\
&=H\left(\,\sum_{i\in I_C}\vM_i \Big| \vY_{\hC},\vM_{S\setminus I_C},\vK_{S\setminus I_C} \right),
\end{split}
\end{align}
where $\widetilde{C} = C \cup C'$, and clearly, $\widetilde{C}$ is a global cut set, namely that $\widetilde{C}$ separates $\rho$ from all the source nodes in $S$. Since the global cut set $\widetilde{C}$ also separates $\ein(\rho)$ from $S$, we see that $\vY_{\ein(\rho)}=\hg_{\ein(\rho)}(\vM_S,\vK_S)$ is a function of $\vY_{\widetilde{C}}=\hg_{\widetilde{C}}(\vM_S,\vK_S)$. Together with the admissibility of the secure network code $\hmbC$, the target function~$f$ must be computable with zero error on the global cut set $\widetilde{C}$, i.e.,
\begin{align}\label{pf-equ7}
0=H\big(f(\vM_S)|\vY_{\widetilde{C}}\big)=H\left(\,\sum_{i\in S}\vM_i \Big| \vY_{\widetilde{C}}\right).
\end{align}
Now, we continue from \eqref{pf-equ10} to obtain
\begin{align}
& H\left(\,\sum_{i\in I_C}\vM_i \Big| \vY_{C},\vM_{S\setminus I_C},\vK_{S\setminus I_C}\right) = H\left(\,\sum_{i\in I_C}\vM_i \Big| \vY_{\widetilde{C}},\vM_{S\setminus I_C},\vK_{S\setminus I_C}\right) \nonumber\\
& = H\left(\,\sum_{i\in I_C}\vM_i \Big| \vY_{\widetilde{C}}, \sum_{i\in S}\vM_i, \vM_{S\setminus I_C},\vK_{S\setminus I_C} \right)\label{eq_Omin_7}\\
& =0,\label{eq_Omin_7-1}
\end{align}
where the equality~\eqref{eq_Omin_7} follows from \eqref{pf-equ7} and the equality~\eqref{eq_Omin_7-1} follows from
\begin{align*}
 H\left(\,\sum_{i\in I_C}\vM_i \Big|\sum_{i\in S}\vM_i, \vM_{S\setminus I_C} \right)=0.
\end{align*}
We thus have proved the equality \eqref{pf-equ5}.

For the cut set $C$, we consider a wiretap set $W \in \mW_r$ such that the pair $(C, W)$ is valid.

\medskip
\noindent\textbf{Case 1:} $D_W = I_C = S$.

In this case, by $I_C = S$ we readily see that $C$ is a global cut set, and thus we rewrite the equality~\eqref{pf-equ5} as $H \big( \sum_{i\in S}\vM_i \Big| \vY_{C} \big)=0$. Together with \eqref{equ1} and the security condition $I\big(\vY_W;\sum_{i \in S}\vM_i\big) = 0$ (cf.~\eqref{eq_sec_condition_for_f}), we have
\begin{align}
\ell\cdot \log q & =  H\left(\,\sum_{i\in S}\vM_i \right) = H\left(\,\sum_{i\in S}\vM_i \Big| \vY_W \right)\nonumber\\
& =  H\left(\,\sum_{i\in S}\vM_i \Big| \vY_W \right)-H\left(\,\sum_{i\in S}\vM_i \Big| \vY_{C}\right)\nonumber\\
& = I\left(\,\sum_{i\in S}\vM_i;\vY_{C \setminus W} \Big| \vY_{W} \right) \nonumber \\
& \leq H\big(\vY_{C \setminus W} \big| \vY_{W} \big) \nonumber \\
& \leq H(\vY_{C \setminus W}) \leq \big| C\setminus W \big| \cdot \log q^n \nonumber \\
&=  n\cdot \big(|C|-|W| \big)\cdot \log q, \nonumber 
\end{align}
i.e.,
\begin{align*}
\frac{\ell }{n}\leq |C|-|W|.
\end{align*}

\noindent\textbf{Case 2:} $I_C \setminus D_W \neq \emptyset$.

In this case, we need the following equality
\begin{equation}\label{eq_sumIC_given_YW}
  H \left(\, \sum_{i\in I_C}\vM_i\right) = H \left(\,\sum_{i\in I_C}\vM_i \Big| \vY_{W},\vM_{S\setminus I_C},\vK_{S\setminus I_C} \right).
\end{equation}
To prove \eqref{eq_sumIC_given_YW}, we consider
\begin{align}
  &H\left(\,\sum_{i \in I_C}\vM_i \Big| \vY_W, \vM_{S\setminus I_C},\vK_{S\setminus I_C}\right)\nonumber\\
  &\geq H\left(\,\sum_{i \in I_C}\vM_i \Big| \vM_{D_W},\vK_{D_W}, \vM_{S\setminus I_C},\vK_{S\setminus I_C}\right) \label{eq_entropy_decrease_S-IC-DW}\\
  &= H\left(\,\sum_{i \in I_C \setminus D_W}\vM_i + \sum_{i \in I_C \cap D_W}\vM_i \Big| \vM_{D_W},\vK_{D_W}, \vM_{S\setminus I_C},\vK_{S\setminus I_C}\right)\nonumber\\
  &= H\left(\,\sum_{i \in I_C \setminus D_W}\vM_i \Big| \vM_{D_W},\vK_{D_W}, \vM_{S\setminus I_C},\vK_{S\setminus I_C}\right)\nonumber\\
  &= H\left(\,\sum_{i \in I_C \setminus D_W}\vM_i \right) = \ell \cdot \log q = H\left(\,\sum_{i \in I_C}\vM_i\right), \label{eq_entropy_remain_information_0-plus}
\end{align}
where the inequality \eqref{eq_entropy_decrease_S-IC-DW} follows from the fact that $\vY_W$ is a function of $(\vM_{D_W}, \vK_{D_W})$; the first equality in~\eqref{eq_entropy_remain_information_0-plus} follows because $(\vM_i:~i \in I_C \setminus D_W)$ and $(\vM_{D_W},\vK_{D_W}, \vM_{S\setminus I_C},\vK_{S\setminus I_C})$ are independent since
\begin{equation*}
  \big(I_C \setminus D_W\big)\cap \big( D_W \cup (S \setminus I_C) \big) = \emptyset;
\end{equation*}
and the second equality in~\eqref{eq_entropy_remain_information_0-plus} follows from $I_C \setminus D_W \neq \emptyset$. Together with the inequality
\begin{equation*}
  H\left(\,\sum_{i \in I_C}\vM_i\right) \geq H\left(\,\sum_{i \in I_C}\vM_i \Big| \vY_W, \vM_{S\setminus I_C},\vK_{S\setminus I_C}\right),
\end{equation*}
we immediately prove the equality \eqref{eq_sumIC_given_YW}.

Now, continuing from \eqref{equ1}, we have
\begin{align}
\ell\cdot \log q & =  H\left(\,\sum_{i\in I_C}\vM_i \right) \nonumber\\
& = H\left(\,\sum_{i\in I_C}\vM_i \Big| \vY_W, \vM_{S\setminus I_C},\vK_{S\setminus I_C}\right)\label{eq_Thm2_plus2}\\
& =  H\left(\,\sum_{i\in I_C}\vM_i \Big| \vY_W, \vM_{S\setminus I_C}, \vK_{S\setminus I_C} \right)-H\left(\,\sum_{i\in I_C}\vM_i \Big| \vY_{C},\vM_{S\setminus I_C},\vK_{S\setminus I_C}\right)\label{eq_Thm2_plus3}\\
& = I\left(\,\sum_{i\in I_C}\vM_i;\vY_{C \setminus W} \Big| \vY_{W},\vM_{S\setminus I_C},\vK_{S\setminus I_C} \right) \nonumber \\
& \leq H\big(\vY_{C \setminus W} \big| \vY_{W},\vM_{S\setminus I_C},\vK_{S\setminus I_C} \big) \nonumber \\
& \leq H(\vY_{C \setminus W}) \leq \big| C\setminus W \big| \cdot \log q^n = n\cdot \big(|C|-|W| \big)\cdot \log q, \nonumber 
\end{align}
i.e.,
\begin{align*}
R(\hmbC)\leq |C|-|W|,
\end{align*}
where the equalities \eqref{eq_Thm2_plus2} and \eqref{eq_Thm2_plus3} follows from \eqref{eq_sumIC_given_YW}and \eqref{pf-equ5}, respectively.

\medskip

Combining the above two cases, we have $\ell / n \leq |C| - |W|$ for all the valid pairs $(C, W) \in \Lambda(\mN) \times \mW_r$. We thus obtain that
\begin{align*}
\frac{\ell }{n} \leq \min_{(C, W)\in \Lambda(\mN) \times \mW_r :\atop (C, W)\,\text{is valid}}  \big( |C|-|W| \big).
\end{align*}
Finally, since the above upper bound is valid for all the admissible secure network codes for the model $\langle \mN, f, r \rangle$, we have proved the upper bound in  \eqref{eq_upper_algebraic_sum}, and hence the theorem is proved.
\end{IEEEproof}

\medskip

The upper bound obtained in Theorem~\ref{thm_general_upper} is applicable to an arbitrary network topology and any security level. Note that this upper bound, which depends on the network topology and security level, is graph-theoretic and not in closed form. Hence, we will give a lower bound and an upper bound on the upper bound in Theorem~\ref{thm_general_upper}, where both of them are in closed form.

\begin{cor}\label{cor_general_upper}
Consider the secure model $\langle \mN,  f, r \rangle$, where the target function $f$ is the algebraic sum over a finite field $\Fq$.

\begin{itemize}
  \item If the security level $r > C_{\min}^S$, then $\hmC \langle \mN, f, r \rangle = 0$.
  \item If the security level $r \leq C_{\min}^S$, the upper bound on $\hmC \langle \mN, f, r \rangle$ obtained in Theorem~\ref{thm_general_upper} satisfies
\begin{align}
C_{\min}-r \leq \min_{(C, W)\in \Lambda(\mN) \times \mW_r :\atop  (C, W)\,\text{\rm is valid}} \big( |C|-|W| \big) \leq \min\big\{C_{\min}, C_{\min}^S - r\big\},\label{eq_upper_lower_algebraic_sum_cor}
\end{align}
\end{itemize}
where
\begin{equation*}
  C_{\min} \triangleq \min_{\sigma \in S}\mincut(\sigma,\rho) \quad \text{and} \quad C_{\min}^S \triangleq \mincut(S, \rho),
\end{equation*}
i.e., $C_{\min}$ is the smallest minimum cut capacity separating the sink node $\rho$ from each source node~$\sigma$ in $S$, and $C_{\min}^S$ is the minimum cut capacity separating $\rho$ from the set of source nodes~$S$.
\end{cor}

\begin{IEEEproof}
For the case that the security level $r > C_{\min}^S$, by \eqref{eq_upper_algebraic_sum} we readily see that
\begin{equation*}
  \hmC \langle \mN, f, r \rangle \leq \min_{(C, W)\in \Lambda(\mN) \times \mW_r :\atop  (C, W)\,\text{is valid}} \big( |C|-|W| \big) = 0,
\end{equation*}
because for any global cut set $C'$ with $|C'|=C_{\min}^S$, it follows from $I_{C'} \setminus D_{C'} = \emptyset$ that the pair~$(C',C')$ is in $\Lambda(\mN) \times \mW_r$ and is valid. Together with the fact that $\hmC \langle \mN, f, r \rangle \geq 0$, we immediately have $\hmC \langle \mN, f, r \rangle = 0$ for this case.

Next, we consider the case that the security level $r \leq C_{\min}^S$. We first prove the lower bound in~\eqref{eq_upper_lower_algebraic_sum_cor}. Consider
\begin{align}
&\min_{(C, W)\in \Lambda(\mN) \times \mW_r :\atop  (C, W)\,\text{is valid}} \big( |C|-|W| \big) \nonumber\\
&~~= \min_{ C \in\Lambda(\mN) } \Big( |C|-\max_{ W \in \mW_r:  \atop (C, W)\,\text{is valid}} |W| \Big) \nonumber\\
&~~\geq C_{\min}-r \label{ineq1-pf-ineq_cor_up-b_low-b},
\end{align}
where the inequality~\eqref{ineq1-pf-ineq_cor_up-b_low-b} follows from $|W|\leq r$ for all $W \in \mW_r$ and $|C| \geq C_{\min}$ for all $C \in \Lambda(\mN)$.

Next, we prove the upper bound in \eqref{eq_upper_lower_algebraic_sum_cor}.

\noindent\textbf{Case 1:} To prove the upper bound $C_{\min}$, we consider a cut set $C_1 \in \Lambda(\mN)$ with $|C_1| = C_{\min}$ and the empty wiretap set $W = \emptyset$. Clearly, $(C_1, W)$ is because $I_{C_1} \setminus D_W = I_{C_1} \neq \emptyset$. This immediately implies that
\begin{equation}\label{eq_bound_less_than_Cmin}
  \min_{(C, W)\in \Lambda(\mN) \times \mW_r :\atop  (C, W)\,\text{is valid}} \big( |C|-|W| \big) \leq |C_1| - |\emptyset| = C_{\min}.
\end{equation}

\noindent\textbf{Case 2:} To prove the upper bound $C_{\min}^S - r$, we consider a global cut set $C_2 \in \Lambda(\mN)$ with $|C_2| = C_{\min}^S$. Then, it follows from $I_{C_2} = S$ that the pair $(C_2, W)$ is valid for each wiretap set $W \in \mW_r$ satisfying $W \subseteq C_2$, because either $D_W=I_{C_2}=S$ or $D_W \subsetneq I_{C_2}=S$, implying that $I_{C_2} \setminus D_W = S \setminus D_W \neq \emptyset$. Thus, we obtain that
\begin{equation}\label{eq_bound_less_than_CminS-r}
  \min_{(C, W)\in \Lambda(\mN) \times \mW_r :\atop  (C, W)\,\text{is valid}} \big( |C|-|W| \big) \leq |C_2|  - \max_{W \in \mW_r: \atop  (C_2, W)\,\text{is valid}} |W| = C_{\min}^S - r,
\end{equation}
where the last equality follows from $|C_2| = C_{\min}^S$ and
\begin{equation*}
  \max_{W \in \mW_r: \atop  (C_2, W)\,\text{is valid}} |W|  = r
\end{equation*}
because $r \leq C_{\min}^S$.

Combining \eqref{eq_bound_less_than_Cmin} and \eqref{eq_bound_less_than_CminS-r}, we have proved the upper bound in \eqref{eq_upper_lower_algebraic_sum_cor}. The corollary is thus proved.
\end{IEEEproof}

\medskip

A straightforward consequence of Corollary~\ref{cor_general_upper} is given below.

\begin{cor}\label{lemma_max_r}
Consider the secure model $\langle \mN,  f, r \rangle$, where the target function $f$ is the algebraic sum over a finite field $\Fq$. If the security level $r = 0$, then
  \begin{equation*}
    \hmC\langle \mN,  f, 0 \rangle = C_{\min}.
  \end{equation*}
\end{cor}

In fact, $C_{\min}$ is the computing capacity for the algebraic sum over an arbitrary network without any security consideration, i.e., $r=0$ \cite{YangGuang-TIT18,Guang_NFC_TIT19, PartI}. Thus, the upper bound~\eqref{eq_upper_algebraic_sum} obtained in Theorem~\ref{thm_general_upper} degenerates to the computing capacity for the algebraic sum $f$ over a network $\mN$. However, we remark that the security level $r=0$ is not necessary for our upper bound to be equal to~$C_{\min}$ (e.g., see Example~\ref{example_toy}).

To end this section, we consider a special type of networks $\mN$ with the topology property $C_{\min} = C_{\min}^S$. For such a network $\mN$, the secure computing capacity $\hmC \langle \mN, f, r \rangle$ can be fully characterized, as stated in the following corollary.

\begin{cor}\label{cor_Cmin=Cgmin}
Consider the secure model $\langle \mN, f, r \rangle$, where $C_{\min} = C_{\min}^S$ for the network $\mN$. Then
\begin{align*}
    \hmC\langle \mN, f, r \rangle = \begin{cases}
                                      C_{\min}^S - r & \mbox{if } r \leq C_{\min}^S; \\
                                      0 & \mbox{if } r > C_{\min}^S.
                                    \end{cases}
\end{align*}
\end{cor}
\begin{IEEEproof}
By Corollary~\ref{lemma_max_r}, we immediately obtain that $\hmC \langle \mN, f, r \rangle = 0$ for the security level $r > C_{\min}^S$. Thus, it suffices to prove that $\hmC \langle \mN, f, r \rangle = C_{\min}^S - r$ for any security level $r \leq C_{\min}^S$.

First, it follows from $C_{\min} = C_{\min}^S$ that for $r \leq C_{\min}^S$, the LHS and RHS of \eqref{eq_upper_lower_algebraic_sum_cor} in Corollary~\ref{cor_general_upper} are equal, implying that
\begin{equation}\label{eq_cor_4_plus1}
    \hmC\langle \mN, f, r \rangle \leq \min_{(C, W)\in \Lambda(\mN) \times \mW_r :\atop  (C, W)\,\text{is valid}} \big( |C|-|W| \big) = C_{\min}^S - r.
\end{equation}
On the other hand, it was proved in \cite[Theorem 10]{PartI} that for the secure model $(\mN, f, r)$ with $r \leq C_{\min}$, where the security of the information sources are considered, i.e., $I\big(\vM_S; \vY_W\big) = 0$ for all $W \in \mW_r$, the secure computing capacity $\hmC(\mN, f, r)$ is lower bounded by $C_{\min} - r$, i.e.,
  \begin{equation}\label{eq_cor_4_plus2}
    \hmC(\mN, f, r) \geq C_{\min} - r.
  \end{equation}
 Further, we note that
   \begin{equation}\label{eq_cor_4_plus3}
     \hmC \langle \mN, f, r \rangle \geq \hmC (\mN, f, r)
   \end{equation}
    because an admissible secure network code for $(\mN, f, r)$ must be admissible for $\langle \mN, f, r \rangle$. Combining~\eqref{eq_cor_4_plus2} and \eqref{eq_cor_4_plus3}, we thus have proved that
    \begin{equation*}
      \hmC \langle \mN, f, r \rangle \geq \hmC (\mN, f, r) \geq C_{\min} - r = C_{\min}^S - r.
    \end{equation*}
    Together with the upper bound \eqref{eq_cor_4_plus1}, we have proved that $\hmC \langle \mN, f, r \rangle = C_{\min}^S - r$, and thus the corollary.
\end{IEEEproof}

\subsection{A Comparison of the Upper Bounds for the Models $\langle \mN, f, r \rangle$ and $(\mN, f, r)$}\label{subsec_comp_bound}

First, we recall the source-security model $(\mN, f, r)$ investigated in Part \Rmnum{1} of the current paper \cite{PartI}, where all the information sources are protected from being leaked to the wiretapper, i.e., $I\big(\vM_S; \vY_W\big) = 0$ for all $W \in \mW_r$. For this model $(\mN, f, r)$, a nontrivial upper bound on the secure computing capacity $\hmC(\mN, f, r)$  was proved, which we state below.

\begin{thm}[\!\!{\cite[Theorem 1]{PartI}}]\label{thm_bound_on_souce}
Consider the secure model $(\mN, f, r)$, where the target function~$f$ is an algebraic sum over a finite field $\Fq$. Then,
\begin{align}\label{eq_bound_on_souce}
   \hmC(\mN, f, r) \leq
   \min_{ (C,W)\in \Lambda(\mN) \times \mW_r:\atop   W\subseteq C \text{ \rm  and }  D_W \subseteq  I_C} \big( |C|-|W| \big).
   \end{align}
\end{thm}

Clearly, the secure computing capacity for the target-function-security model $\langle \mN, f, r \rangle$ is lower bounded by the secure computing capacity for the source-security model $(\mN, f, r)$, i.e., \eqref{eq_cor_4_plus3}. The following theorem asserts that the upper bound on $\hmC \langle \mN, f, r \rangle$ obtained in Theorem~\ref{thm_general_upper} is also lower bounded by the above upper bound \eqref{eq_bound_on_souce} on $\hmC(\mN, f, r)$.

\begin{thm}
The two upper bounds on the secure computing capacities $\hmC \langle \mN, f, r \rangle$ and $\hmC(\mN, f, r)$ presented in Theorem~\ref{thm_general_upper} and Theorem~\ref{thm_bound_on_souce}, respectively, satisfy
\begin{equation}\label{eq_size_relationship}
  \min_{ (C, W)\in \Lambda(\mN) \times \mW_r :\atop  (C, W)\,\text{\rm is valid}} \big( |C|-|W| \big) \geq \min_{ (C, W)\in \Lambda(\mN) \times \mW_r :\atop  W\subseteq C\,\text{\rm and}\,D_W \subseteq I_C} \big( |C|-|W| \big).
\end{equation}
\end{thm}

\begin{IEEEproof}
Let $(C, W) \in \Lambda(\mN) \times \mW_r$ be an any valid pair. Then, the pair $(C, W)$ satisfies either \textbf{cond~1} or \textbf{cond~2}. Now, we write the upper bound on $\hmC \langle \mN, f, r \rangle$ obtained in Theorem~\ref{thm_general_upper} (i.e., the left hand side of \eqref{eq_size_relationship}) as follows:
  \begin{equation}\label{eq_min_separate}
    \min_{ (C, W)\in \Lambda(\mN) \times \mW_r :\atop  (C, W)\,\text{\rm is valid}} \big( |C|-|W| \big) = \min\left\{\min_{ (C, W)\in \Lambda(\mN) \times \mW_r :\atop  (C, W)\,\text{\rm satisfies \textbf{cond 1}}} \big( |C|-|W| \big),~\min_{ (C, W)\in \Lambda(\mN) \times \mW_r :\atop  (C, W)\,\text{\rm satisfies \textbf{cond 2}}} \big( |C|-|W| \big)\right\}.
  \end{equation}

  First, we readily see that
    \begin{equation}\label{eq_minCase2}
    \min_{ (C, W)\in \Lambda(\mN) \times \mW_r :\atop  (C, W)\,\text{\rm satisfies \textbf{cond 2}}} \big( |C|-|W| \big) \geq \min_{ (C, W)\in \Lambda(\mN) \times \mW_r :\atop  W\subseteq C~\text{and}~D_W \subseteq I_C} \big( |C|-|W| \big),
  \end{equation}
  because $W \subseteq C$ and $D_W \subseteq I_C$ for any $(C, W) \in \Lambda(\mN) \times \mW_r$ that satisfies \textbf{cond 2}.

  Next, we will prove that
\begin{equation}\label{eq_minCase1}
    \min_{ (C, W)\in \Lambda(\mN) \times \mW_r :\atop  (C, W)\,\text{\rm satisfies \textbf{cond 1}}} \big( |C|-|W| \big) \geq \min_{ (C, W)\in \Lambda(\mN) \times \mW_r :\atop  W\subseteq C~\text{and}~D_W \subseteq I_C} \big( |C|-|W| \big).
  \end{equation}
  Let $(C^*, W^*) \in \Lambda(\mN) \times \mW_r$ be a pair that satisfies \textbf{cond 1} such that
    \begin{equation*}
     |C^* \setminus W^*| = |C^*| - |W^*| = \min_{ (C, W)\in \Lambda(\mN) \times \mW_r :\atop  (C, W)\,\text{\rm satisfies \textbf{cond 1}}} \big( |C|-|W| \big).
   \end{equation*}
Now, for the pair $(C^*, W^*)$, we claim that
   \begin{equation}\label{eq_thm6_plus1}
       I_{C^*} \setminus D_{W^*} \subseteq I_{C^* \setminus W^*},
   \end{equation}
   which will be proved later. Together with $I_{C^*} \setminus D_{W^*} \neq \emptyset$,  this immediately implies that $C^* \setminus W^* \in \Lambda(\mN)$. Thus
   \begin{equation*}
     (C^* \setminus W^*, \emptyset) \in \Big\{(C, W) \in \Lambda(\mN) \times \mW_r: W \subseteq C~\text{and}~D_W \subseteq I_C \Big\},
   \end{equation*}
and it follows immediately that
   \begin{equation*}
     \min_{ (C, W)\in \Lambda(\mN) \times \mW_r :\atop  (C, W)\,\text{\rm satisfies \textbf{cond 1}}} \big( |C|-|W| \big) = |C^* \setminus W^*| \geq \min_{ (C, W)\in \Lambda(\mN) \times \mW_r :\atop  W\subseteq C~\text{and}~D_W \subseteq I_C} \big( |C|-|W| \big),
   \end{equation*}
   i.e., the inequality \eqref{eq_minCase1}. Combining \eqref{eq_minCase2} and \eqref{eq_minCase1} with \eqref{eq_min_separate}, we have proved \eqref{eq_size_relationship}.

Now, we prove \eqref{eq_thm6_plus1}. We assume the contrary that $I_{C^*} \setminus D_{W^*} \not \subseteq I_{C^* \setminus W^*}$, i.e., there exists a source node $\sigma \in I_{C^*} \setminus D_{W^*}$ but $\sigma \notin I_{C^* \setminus W^*}$. This implies that \rmnum{1})~by $\sigma \notin D_{W^*}$, every path from~$\sigma$ to $\rho$ does not pass through any edge in $W^*$; \rmnum{2})~by $\sigma \notin I_{C^* \setminus W^*}$, there exists a path, say $P$, from $\sigma$ to $\rho$ such that $P$ does not pass through any edge in $C^* \setminus W^*$. It follows from \rmnum{1})~and \rmnum{2})~that the path~$P$ from~$\sigma$ to $\rho$ does not pass through any edge in $C^*$, a contradiction to the fact that $\sigma \in I_{C^*}$. We have thus proved \eqref{eq_thm6_plus1}. Hence, the theorem is proved.
\end{IEEEproof}

\section{Linear (Function-Computing) Secure Network Coding}\label{subsec_LNC}

For two positive integers $\ell$ and $n$, an $(\ell, n)$ (function-computing) secure network code for the model $\langle\mN, f, r\rangle$ is said to be \emph{linear} if the local encoding function for each edge is linear. This definition for linearity of the secure network code is the same as that for the model $(\mN, f, r)$ in \cite{PartI}. In particular, when $n=1$, an $(\ell, n)$ linear secure network code is also referred to as a \emph{scalar linear secure network code}.\footnote{Such an $(\ell,n)$ linear secure network code with $n>1$ sometimes is also called ``vector-linear''.}

We now elaborate the definition of scalar linear secure network codes, which is of special importance in the rest of the paper. For a positive integer $\ell$, each source node $\sigma_i$ sequentially generates~$\ell$ i.i.d. random variables $M_{i,1},$ $M_{i,2},\cdots, M_{i,\ell}$ with generic random variable $M_i$. We let $\vM_i=(M_{i,1},M_{i,2}, \cdots, M_{i,\ell})$ and $\vM_S=(\vM_1,\vM_2,\cdots,\vM_s)$. Furthermore, we recall that $\vK_i$ denotes the random key available at the source node $\sigma_i$ which is now assumed to be distributed uniformly on the vector space $\Fq^{z_i}$, where $z_i$ is a nonnegative integer.\footnote{When $z_i = 0$, we have $\Fq^{0} = \{\emptyset\}$.} Let $\vK_S=(\vK_1,\vK_2,\cdots,\vK_s)$. Now, for $i=1,2,\cdots,s$, we let $\vm_i \in \Fq^{\ell}$ and $\vk_i\in \Fq^{z_i}$ be arbitrary outputs of the source message~$\vM_i$ and the key~$\vK_i$, respectively. Accordingly, let $\vm_S = (\vm_1,\vm_2,\cdots,\vm_s)$ and $\vk_S = (\vk_1,\vk_2,\cdots,\vk_s)$, i.e., $\vm_S$ and $\vk_S$ are arbitrary values taken by $\vM_S$ and~$\vK_S$, respectively. Further, we let $\vx_i=(\vm_i~\vk_i)$ for $i=1,2,\cdots,s$, and let $\vx_S=(\vx_1~\vx_2~ \cdots~\vx_s)$.

An $(\ell,1)$ (scalar) linear secure network code $\hmbC=\big\{ \htheta_{e}:~e\in \mE;~  \hvarphi \big\}$ consists of a {\em linear} local encoding function~$\htheta_e$ for each edge~$e$ and a decoding function $\hvarphi$ at the sink node $\rho$ as follows:

\begin{itemize}
  \item For each linear local encoding function $\htheta_e$ associated with edge $e$,
\begin{align*}
    \htheta_e:
    \begin{cases}
      \Fq^{\ell} \times \Fq^{z_i} \rightarrow \Fq & \mbox{if } \tail(e) = \sigma_i \mbox{ for some } i;\\
      \prod\limits_{d \in \ein(\tail(e))}\Fq \rightarrow \Fq & \mbox{otherwise},
    \end{cases}
  \end{align*}
  and more precisely,
  \begin{align}\label{def_linear_local}
    \begin{cases}
      \htheta_e(\vm_i~\vk_i) = (\vm_i~\vk_i) \cdot A_{i, e} & \mbox{if } \tail(e) = \sigma_i \mbox{ for some } i;\\
      \htheta_e\big(y_d: d \in \ein(\tail(e))\big) = \sum\limits_{d \in \ein(\tail(e))}a_{d, e} \cdot y_d & \mbox{otherwise},
    \end{cases}
  \end{align}
  where $A_{i, e}$ is an $\Fq$-valued column $(\ell+z_i)$-vector, $y_d \in \Fq$ is the message transmitted on an edge~$d$, and $a_{d, e} \in \Fq$ is the \emph{local encoding coefficient} of the adjacent edge pair $(d, e)$ for the code $\hmbC$;

  \item The decoding function $\hvarphi$ at the sink node $\rho$ is a mapping from $\prod_{\ein(\rho)}\Fq$ to $\Fq^\ell$ for computing the algebraic sum $f$ with zero error.
\end{itemize}

With the causality of the linear encoding scheme described in \eqref{def_linear_local}, we see that $y_e$ for each edge $e \in \mE$ is a linear function of $\vm_S$ and $\vk_S$, or equivalently, the global encoding function $\hg_e$ for each $e\in \mE$, induced by the linear local encoding functions, is linear. Thus, for each $e\in \mE$, there exists an $\Fq$-valued column vector $\vg_e$ with dimension $\ell s+\sum_{i=1}^{s}z_i$ such that
\begin{align}
\hg_{e}\big(\vm_S~\vk_S\big) = \big((\vm_1~\vk_1)~~(\vm_2~\vk_2)~~\cdots~~(\vm_s~\vk_s)\big) \cdot \vg_e = y_e,\label{eq_linear_global_vector}
\end{align}
where for notational simplicity, we follow~\cite{PartI} to use $(\vm_S~\vk_S)$ to represent $\big((\vm_1~\vk_1)~~(\vm_2~\vk_2)~~\cdots~~(\vm_s~\vk_s)\big)$  in spite of an abuse of notation, and call $\vg_e$ the \emph{global encoding vector} of the edge $e$. Furthermore, we write
\begin{align}\label{eq_ge_separate_sigma}
\vg_e=\begin{bmatrix}\vg_e^{\,(\sigma_1)} \\ \vg_e^{\,(\sigma_2)} \\ \vdots \\ \vg_e^{\,(\sigma_s)}\end{bmatrix}
~~\text{with}~~\vg_e^{\,(\sigma_i)} \in \Fq^{\ell+z_i},~\forall~1\leq i \leq s.
\end{align}
Then the second equality \eqref{eq_linear_global_vector} can be rewritten as
\begin{align*}
  \sum_{i=1}^{s} (\vm_i~\vk_i) \cdot \vg_e^{\,(\sigma_i)} = y_e,~~\forall~e \in \mE.
\end{align*}
Also, for an edge subset $W \subseteq \mE$, we let $G_W \triangleq \Big[\vg_e: e \in W\Big]$ and write
\begin{align*}
G_W=\begin{bmatrix}G_W^{\,(\sigma_1)} \\ G_W^{\,(\sigma_2)} \\ \vdots \\ G_W^{\,(\sigma_s)}\end{bmatrix},
\end{align*}
where $G_W^{\,(\sigma_i)} \triangleq \Big[\vg^{\,(\sigma_i)}_e:~e \in W\Big]$ is a matrix of size $(\ell+z_i) \times |W|,~1 \leq i \leq s$.

Next, for such an $(\ell, 1)$ linear secure network code $\hmbC$ for the model $\langle \mN, f, r \rangle$, we define a matrix $T(\hmbC)$ as follows:
\begin{align}\label{eq_GammaC_def}
    \matT(\hmbC) \triangleq
    \begin{bmatrix}
      \matT^{(1)} \\\smallskip
      \matT^{(2)} \\\smallskip
      \vdots \\\smallskip
     \matT^{(s)}
    \end{bmatrix}
    ~\text{ with }~
    \matT^{(i)} =
    \begin{bmatrix}
      I_{\ell} \\
      \mbO_{z_i \times \ell}
    \end{bmatrix},~\forall~1\leq i\leq s,
  \end{align}
  where $I_\ell$ is the $\ell \times \ell$ identity matrix and $\mbO_{z_i \times \ell}$ is a $z_i \times \ell$ all-0 matrix.\footnote{In the rest of the paper, we simply write $\mbO_{z_i \times \ell}$ as $\mbO$ if the dimensions are clear from the context.} Evidently, the matrix $\matT(\hmbC)$ is of size $\big(\sum_{i=1}^{s}(\ell+z_i)\big) \times \ell$ and in the rest of the paper we write $\matT(\hmbC)$ as $\matT$ for notational simplicity when there is no ambiguity on the code $\hmbC$. With this, we can see that
  \begin{align}
  \big(\vM_S~\vK_S\big) \cdot\matT =\sum_{i=1}^{s}\vM_i = f(\vM_S).\label{eq_XSGA=fMS}
  \end{align}
  It further follows from the target-function-security condition \eqref{eq_sec_condition_for_f} for the model $\langle \mN, f, r \rangle$ that
\begin{align}\label{eq_equivalent_function_security_condition}
        I\big(\vY_W; f(\vM_S)) = I\big(\vY_W;\big(\vM_S~\vK_S\big) \cdot\matT\big) = 0,~~\forall~W \in \mW_r.
\end{align}

The following theorem provides an equivalent target-function-security condition.

\begin{thm}\label{thm_sec_condition_space_general}
  Consider the secure model $\langle \mN, f, r \rangle$ with $f$ being the algebraic sum over $\Fq$. Let $\hmbC$ be an $(\ell, 1)$ linear secure network code, of which the global encoding vectors are $\vg_e,~e\in\mE$. Then, the target-function-security condition \eqref{eq_equivalent_function_security_condition} is satisfied for the code $\hmbC$ if and only if
  \begin{align}\label{def_sec_condition_space}
  \big\langle G_W \big\rangle \cap \big\langle \matT \big\rangle = \big\{\vec{0}\big\},~~\forall~W \in \mW_r. \footnotemark
  \end{align}
\end{thm}
\footnotetext{Here, we use $\big\langle \mat  \big\rangle$ to denote the linear space spanned by the column vectors of a matrix $\mat$.}

\begin{IEEEproof}
  We first prove the ``\,only if\,'' part by contradiction. Suppose the contrary that there exists a wiretap set $W \in \mW_r$ that does not satisfy the condition \eqref{def_sec_condition_space}, i.e.,
  \begin{align}\label{def_sec_condition_space_n0}
    \big\langle G_W  \big\rangle \cap \big\langle \matT \big\rangle \neq \big\{\vec{0}\big\}.
  \end{align}
  We will prove that $I\big(f(\vM_S) ;~\vY_W\big) > 0$. By \eqref{def_sec_condition_space_n0}, there exist two non-zero column vectors $\vec{\alpha} \in \Fq^{|W|}$ and $\vec{\beta} \in \Fq^{\ell}$ such that
  \begin{align}
    G_W \cdot \vec{\alpha}= \matT \cdot \vec{\beta} \neq \vec{0}.\label{eq_HWw=Gav}
  \end{align}
  We now consider
\begin{align}
        &I\big(f(\vM_S); \vY_W\big) \nonumber\\
        &=I\big(f(\vM_S); (\vM_S~\vK_S) \cdot G_W\big)\nonumber\\
        &=H\big(f(\vM_S)\big)-H\big(f(\vM_S) \,\big|\, (\vM_S~\vK_S)  \cdot G_W\big)\nonumber\\
        &=H\big(f(\vM_S)\big) - H\big(f(\vM_S) \,\big|\, (\vM_S~\vK_S) \cdot G_W, (\vM_S~\vK_S) \cdot G_W \cdot \vec{\alpha}\big)\nonumber\\
       & \geq H\big(f(\vM_S)\big)-H\big(f(\vM_S) \,\big|\, (\vM_S~\vK_S) \cdot G_W \cdot \vec{\alpha}\big)\nonumber\\
        &=I\big(f(\vM_S); (\vM_S~\vK_S) \cdot G_W \cdot \vec{\alpha}\big)\nonumber\\
        &= I\big(f(\vM_S); (\vM_S~\vK_S) \cdot \matT \cdot \vec{\beta} \,\big) \label{eq_Gav}\\
        & =I\big(f(\vM_S); f(\vM_S) \cdot \vec{\beta}\, \big)\label{eq_HMLd*=0}\\
        &=H\big(f(\vM_S) \cdot \vec{\beta}\, \big) - H\big(f(\vM_S) \cdot \vec{\beta} \,\big|\, f(\vM_S)\big) \nonumber\\
        &= H\big(f(\vM_S) \cdot \vec{\beta}\, \big) \nonumber\\
        &> 0, \label{eq_If_fb>0}
\end{align}
  where \eqref{eq_Gav} follows from \eqref{eq_HWw=Gav}, \eqref{eq_HMLd*=0} follows from \eqref{eq_XSGA=fMS}, and \eqref{eq_If_fb>0} follows from $\vec{\beta} \neq 0$. Then, we have thus proved that $I\big(f(\vM_S) ;~\vY_W\big) > 0$, a contradiction to the target-function-security condition \eqref{eq_equivalent_function_security_condition}.

  Next, we prove the ``\,if\,'' part, namely that if the condition \eqref{def_sec_condition_space} is satisfied, then the target-function-security condition \eqref{eq_equivalent_function_security_condition} is satisfied, or equivalently,
  \begin{align*}
    H\big(f(\vM_S) \big| \vY_W\big) = H\big(f(\vM_S)\big),~~\forall~W \in \mW_r.
  \end{align*}
  Toward this end, it suffices to prove that for each $W \in \mW_r$, the equality
  \begin{align}
    \Pr\big(f(\vM_S) = \vf \big| \vY_W = \vy\big) =  \Pr\big(f(\vM_S) = \vf\big)\label{eq_x_given_y=x}
  \end{align}
  holds for any row vector $\vf \in \Fq^{\ell}$ and any row vector $\vy \in \Fq^{|W|}$ with $\Pr\big(\vY_W = \vy\big) >0$, where the latter means that  there exists a pair~$(\vm_S~\vk_S)$ of a vector of source messages $\vm_S$ and a vector of keys $\vk_S$ such that
  \begin{align*}
  (\vm_S~\vk_S) \cdot \Big[\vg_e:~e\in W\Big] = \vy.
  \end{align*}

Recall that $M_{i, j},~1\leq i \leq s,~1\leq j \leq \ell$ are i.i.d. random variables according to the uniform distribution on $\Fq$. Then, the $\ell$ algebraic sums $\sum_{i=1}^{s}M_{i, j},~1\leq j \leq \ell$ are also i.i.d. random variables uniformly distributed on $\Fq$. Therefore, $f(\vM_S)$ is uniformly distributed on $\Fq^\ell$, i.e.,
\begin{align}\label{eq_Prf=uniform}
  \Pr \big( f(\vM_S) = \vf) = \frac{1}{q^\ell}, \quad\forall~\vf \in \Fq^\ell.
\end{align}
Next, we consider
\begin{align}
  & \Pr\big( f(\vM_S) = \vf \big| \vY_W = \vy \big) \nonumber\\
 & = \frac{\Pr\big(f(\vM_S) = \vf, \vY_W = \vy\big)}{\Pr\big( \vY_W = \vy\big)} \nonumber\\
 & =\frac{\Pr\big((\vM_S~\vK_S) \cdot \matT = \vf,~ (\vM_S~\vK_S) \cdot G_W = \vy\big)}{\Pr\big((\vM_S~\vK_S) \cdot G_W = \vy\big)}\nonumber\\
  &=\frac{\Pr\big((\vM_S~\vK_S) \cdot \big[\matT~G_W\big] = (\vf~\vy)\big)}{\Pr\big((\vM_S~\vK_S) \cdot G_W = \vy\big)}\nonumber\\
  &=\frac{\sum_{(\vm_S\,\vk_S): (\vm_S\,\vk_S) \cdot [\matT~G_W] = (\vf\,\vy)}\Pr\big(\vM_S=\vm_S,\vK_S=\vk_S) \big)}{\sum_{(\vm_S'\,\vk_S'): (\vm_S'\,\vk_S') \cdot G_W = \vy} \Pr\big(\vM_S=\vm_S', \vK_S=\vk_S')\big)}\nonumber\\
  &= \frac{\#\big\{(\vm_S~\vk_S): (\vm_S~\vk_S) \cdot \big[\matT~G_W\big] = (\vf~\vy)\big\}}
                  {\# \big\{(\vm_S'~\vk_S'): (\vm_S'~\vk_S') \cdot G_W = \vy\big\}} \label{eq_sec_cond_prob_2},
\end{align}
where in \eqref{eq_sec_cond_prob_2}, we use ``$\#\{\cdot\}$'' to denote the cardinality of the set and the equality holds because $\vM_S$ and $\vK_S$ are independent and uniformly distributed on $\big(\Fq^\ell\big)^s$ and $\prod_{i=1}^{s}\Fq^{z_i}$, respectively.

\begin{itemize}
  \item For the dominator of \eqref{eq_sec_cond_prob_2}, we have
\begin{align}
  \#\Big\{(\vm_S'~\vk_S'):~ (\vm_S'~\vk_S') \cdot G_W = \vy\Big\} = q^{(s\ell + \sum_{i=1}^{s}z_i) - \Rank(G_W)}.\label{eq_RGW}
\end{align}
  \item For the numerator of \eqref{eq_sec_cond_prob_2}, we have
  \begin{align}
    &\#\Big\{(\vm_S~\vk_S): ~(\vm_S~\vk_S) \cdot \Big[\matT~G_W\Big] = (\vf~\vy)\Big\}\nonumber\\
    &=q^{(s\ell + \sum_{i=1}^{s}z_i) - \Rank([\matT~G_W])} \nonumber\\
    &=q^{(s\ell + \sum_{i=1}^{s}z_i) - \Rank(\matT) - \Rank(G_W)}\label{eq_RGammaGW_new}\\
    &=q^{(s\ell + \sum_{i=1}^{s}z_i) - \ell - \Rank(G_W)},\label{eq_RGammaGW}
  \end{align}
where \eqref{eq_RGammaGW_new} follows from \eqref{def_sec_condition_space} and \eqref{eq_RGammaGW} follows from $\Rank(\matT) = \ell$ by \eqref{eq_GammaC_def}.
\end{itemize}
Finally, combining \eqref{eq_Prf=uniform}, \eqref{eq_sec_cond_prob_2}, \eqref{eq_RGW} and \eqref{eq_RGammaGW}, we obtain that
\begin{align*}
  \Pr\big(f(\vM_S) = \vf~\big|\vY_W = \vy\big) = \frac{1}{q^{\ell}} = \Pr\big(f(\vM_S) = \vf\big),
\end{align*}
i.e., \eqref{eq_x_given_y=x}. Hence, the theorem is proved.
\end{IEEEproof}

\section{Construction of Linear Secure Network Codes for Target-Function Security}\label{sec_code_construction}

We consider the secure model $\langle \mN, f, r\rangle$ with security level $0 \leq r \leq C_{\min}$, where we recall that $C_{\min} \triangleq  \min_{1\leq i \leq s}\mincut(\sigma_i, \rho)$. For any nonnegative integer~$R$ with $r \leq R \leq C_{\min}$, we can construct an admissible $(R-r, 1)$ linear secure network code for the model $\langle \mN,f,r\rangle$. This code construction immediately implies the lower bound $C_{\min}-r$ on the secure computing capacity $\hmC\langle \mN,f,r\rangle$.

\subsection{Code Construction}\label{subsec_code_construction}

Before presenting the code construction, we introduce the following \emph{code transformation}. Consider the model of computing the algebraic sum~$f$ over the network $\mN$ without any security consideration. Denote this model by $(\mN, f)$, and we can construct an admissible $(R, 1)$ linear network code $\mbC$ for $(\mN, f)$,\footnote{Here, we say a linear network code $\mbC$ for $(\mN, f)$ is admissible if the target function $f$ can be computed at the sink node with zero error, i.e., only the computability condition is satisfied.} where the global encoding vectors are $\vh_e \in \Fq^R$ for all $e \in \mE$ (see \cite{PartI} for details). Based on the admissible $(R, 1)$ linear network code $\mbC$, in the following we will construct an admissible $(R-r, 1)$ linear secure network code $\hmbC$ for the secure model $\langle \mN, f, r \rangle$ by using a code transformation to be discussed. Here, $R$ and $r$ are arbitrary nonnegative integers with $r \leq R \leq C_{\min}$.

First, for the code $\mbC$, similar to \eqref{eq_ge_separate_sigma}, for each $e \in \mE$ we write
\begin{equation}\label{eq_he}
  \vh_e = \begin{bmatrix} \vh_e^{(\sigma_1)} \\ \vh_e^{(\sigma_2)}\\\vdots\\\vh_e^{(\sigma_s)} \end{bmatrix}
  ~~\text{with}~~\vh_e^{\,(\sigma_i)} \in \Fq^{R},~\forall~1\leq i \leq s.
\end{equation}
Further, we let $B_i$ be an $R \times R$ invertible matrix over the field $\Fq$,  called the \emph{transformation matrix} (at the source node $\sigma_i$) for $i=1, 2, \cdots, s$, and accordingly, let
\begin{align*}
\widehat{B} = \Big[ B_1~~B_2~~\cdots~~B_s  \Big]_s^{\text{diagonal}}
\triangleq
\begin{bmatrix}
B_1 & \bfzero & \cdots & \bfzero &\\
\bfzero & B_2 & \cdots & \bfzero &\\
\cdots  & \cdots & \cdots & \cdots\\
\bfzero & \bfzero & \cdots & B_s\\
\end{bmatrix}_{Rs \times Rs},
\end{align*}
an $s\times s$ block matrix in which all the blocks in the diagonal are $B_1, B_2, \cdots, B_s$ and all the other blocks are the $R \times R$ zero matrix. Now, we define the transformation of the code $\mbC$ by the matrix $\widehat{B}$, denoted by $\hmbC \triangleq \widehat{B} \cdot\mbC$, which is also an admissible $(R, 1)$ linear network code for $(\mN, f)$ with all the global encoding vectors being $\vg_e \triangleq \widehat{B} \cdot \vh_e,~e \in\mE$
(cf.~\cite{PartI}). Further, for each global encoding vector $\vg_e$ of $\hmbC$, we have
\begin{align*}
\vg_e=\widehat{B} \cdot \vh_e = \begin{bmatrix} B_1 \cdot\vh_e^{\,(\sigma_1)} \\ B_2 \cdot \vh_e^{\,(\sigma_2)}\\\vdots\\B_s \cdot \vh_e^{\,(\sigma_s)}\end{bmatrix}.
\end{align*}
Similar to \eqref{eq_ge_separate_sigma}, we write
\begin{equation*}
  \vg_e = \begin{bmatrix} \vg_e^{\,(\sigma_1)} \\ \vg_e^{\,(\sigma_2)}\\\vdots\\\vg_e^{\,(\sigma_s)} \end{bmatrix}~~\text{with}~~\vg_e^{\,(\sigma_i)} = B_i \cdot \vh_e^{\,(\sigma_i)} \in \Fq^{R}, ~\forall ~1\leq i \leq s.
\end{equation*}

Next, we will design appropriate matrices $B_1, B_2, \cdots, B_s$ such that $\hmbC = \widehat{B} \cdot \mbC$ is an admissible $(R-r, 1)$ linear secure network code for $\langle \mN, f, r \rangle$. First, following \eqref{eq_GammaC_def}, we define a matrix
\begin{align}\label{eq_T_in_construct}
    \matT =
    \begin{bmatrix}
      \matT^{(1)} \\\smallskip
      \matT^{(2)} \\\smallskip
      \vdots \\\smallskip
     \matT^{(s)}
    \end{bmatrix}
    ~\text{ with }~
    \matT^{(i)} =
    \begin{bmatrix}
      I_{R-r} \\
      \mbO_{r \times (R-r)}
    \end{bmatrix},~\forall~1\leq i\leq s.
  \end{align}
Further, we let $B_1, B_2, \cdots, B_s$ be $s$ invertible matrices of size $R \times R$ satisfying the conditions
  \begin{equation}\label{eq_bR_in_Grho}
    \big\langle \widehat{B}^{-1} \cdot \matT  \big\rangle \subseteq \big\langle H_\rho \big\rangle~~\text{with}~~H_\rho \triangleq \begin{bmatrix}
\vh_e:~e \in \ein(\rho)
\end{bmatrix}
\end{equation}
and
\begin{equation}\label{eq_bR_cap_G=0}
    \big\langle \widehat{B}^{-1} \cdot \matT  \big\rangle ~\bigcap~   \big\langle  H_W \big\rangle = \{\vec{0}\}~~\text{with}~~H_W = \Big[\vh_e:~e \in W\Big],~~\forall~W \in \mW_r,
  \end{equation}
which are associated with the computability condition and target-function-security condition, respectively.

\bigskip

In the following, we will prove that $\hmbC \triangleq \widehat{B} \cdot \mbC$, the transformation of the code $\mbC$ by the matrix $\widehat{B}$, is an admissible $(R-r, 1)$ linear secure network code for the secure model $\langle \mN, f, r \rangle$. The existence of the~$s$ transformation matrices $B_1, B_2, \cdots, B_s$ satisfying the conditions \eqref{eq_bR_in_Grho} and \eqref{eq_bR_cap_G=0} will be discussed later. First, for each $i=1,2,\cdots,s$, we recall that the information source $M_i$ generated by the source node~$\sigma_i$ is distributed uniformly on the finite field $\Fq$. Let $\vM_i=(M_{i,1},M_{i,2}, \cdots, M_{i,R-r})$ be the vector of $R-r$ i.i.d. random variables generated by $\sigma_i$ with the generic random variable $M_i$, and then let $\vM_S=(\vM_1,\vM_2,\cdots,\vM_s)$. Furthermore, we let $K_i$ also be a random variable distributed uniformly on~$\Fq$. Let the key available at the source node~$\sigma_i$ be $\vK_i = (K_{i, 1}, K_{i, 2}, \cdots, K_{i, r})$, the vector of $r$ i.i.d. random variables with the generic random variable~$K_i$. Then, the key $\vK_i$ can be regarded as a random variable distributed uniformly on the vector space $\Fq^r$. Let $\vK_S=(\vK_1,\vK_2,\cdots,\vK_s)$. All the keys $\vK_i$ and the source messages $\vM_i$, $i=1,2,\cdots, s$ are mutually independent.
Let $\vm_i \in \Fq^{R-r}$ and $\vk_i\in \Fq^{r}$ be arbitrary valued taken by the source message $\vM_i$ and the key $\vK_i$, respectively, for $i=1,2,\cdots,s$. Accordingly, let $\vm_S = (\vm_1,\vm_2,\cdots,\vm_s)$ and $\vk_S = (\vk_1,\vk_2,\cdots,\vk_s)$, and let $\vx_i=(\vm_i~\vk_i)\in \Fq^R$ and $\vx_S=(\vx_1~\vx_2~ \cdots~\vx_s)$.

\bigskip

\noindent\textbf{Verification of the Computability Condition:}

To implement the code $\hmbC$, i.e., transmitting the message $y_e = \vx_S \cdot \vg_e$ on each edge $e \in \mE$, it is equivalent to linearly transforming $\vx_i$ into $\vx_i \cdot B_i$ at each source node $\sigma_i$, and then using the code $\mbC$ to transmit the message
\begin{equation*}
  (\vx_1\!\cdot\! B_1, ~\vx_2\!\cdot\! B_2,~ \cdots,~ \vx_s\!\cdot\! B_s) \cdot \vh_e = \vx_S \cdot \widehat{B} \cdot \vh_e =\vx_S  \cdot \vg_e = y_e
\end{equation*}
 on each $e \in \mE$. In particular, at the sink node $\rho$, the vector of received messages is
 \begin{equation*}
  \vy_{\rho} \triangleq \big( y_e = \vx_S \cdot \vg_e:~e \in \ein(\rho)\big) = \vx_S \cdot G_\rho,
 \end{equation*}
where
\begin{align*}
G_{\rho} \triangleq \begin{bmatrix}
\vg_e:~e \in \ein(\rho)
\end{bmatrix} = \Big[\widehat{B}\cdot \vh_e:~e\in \ein(\rho)\Big] = \widehat{B}\cdot H_\rho,
\end{align*}
recalling from \eqref{eq_bR_in_Grho} that $H_{\rho} = \begin{bmatrix}
\vh_e:~e \in \ein(\rho)
\end{bmatrix}$.

By \eqref{eq_bR_in_Grho}, there exists a matrix $D$ such that $\widehat{B}^{-1} \cdot \matT = H_\rho \cdot D$. This implies that
\begin{equation*}
  \matT = \widehat{B} \cdot H_\rho \cdot D = G_\rho \cdot D,
\end{equation*}
and thus
\begin{align*}
   \vy_{\rho} \cdot D &= \vx_S \cdot G_\rho \cdot D = \vx_S \cdot \matT = \sum_{i=1}^{s}\vm_i  = \Big( \sum_{i=1}^{s} m_{i, j},~j=1, 2, \cdots, R-r \Big).
\end{align*}
Hence, we have verified the computability condition of the code $\hmbC$.

\bigskip

\noindent\textbf{Verification of the Target-Function-Security Condition:}

Consider an arbitrary wiretap set $W \in \mW_r$. By~\eqref{eq_bR_cap_G=0}, namely that
$\big\langle H_W \big\rangle \bigcap \big\langle \widehat{B}^{-1} \cdot \matT \big\rangle = \{\vec{0}\}$,
we have
\begin{equation}\label{equ_1}
\big\langle \widehat{B} \cdot H_W \big\rangle \cap \big\langle \matT \big\rangle = \{\vec{0}\},
\end{equation}
because otherwise there exist two non-zero column vectors $\vec{\alpha} \in \Fq^{|W|}$ and $\vec{\beta} \in \Fq^{\ell}$ such that
  \begin{align*}
    \widehat{B} \cdot H_W  \cdot \vec{\alpha}= \widehat{B} \cdot \widehat{B}^{-1} \cdot \matT \cdot \vec{\beta} \neq \vec{0},
  \end{align*}
implying that $H_W  \cdot \vec{\alpha} = \widehat{B}^{-1} \cdot \matT \cdot \vec{\beta} \neq \vec{0}$ by the invertibility of $\widehat{B}$, which is a contradiction to \eqref{eq_bR_cap_G=0}. Together with $G_W=\widehat{B} \cdot H_W$, we can rewrite \eqref{equ_1} as
$\big\langle G_W \big\rangle \cap \big\langle \matT \big\rangle = \{\vec{0}\}$. By Theorem~\ref{thm_sec_condition_space_general}, we have already verified the target-function-security condition \eqref{eq_sec_condition_for_f}.

\bigskip

We have verified the computability and target-function-security conditions for the code $\hmbC$. To complete our code construction, it remains to construct $R \times R$ invertible matrices $B_1, B_2, \cdots, B_s$ that satisfy both the conditions~\eqref{eq_bR_in_Grho} and \eqref{eq_bR_cap_G=0}.
First, for each $1 \leq j \leq R-r$, we let $\vec{b}_j$ be an $\Fq$-valued column $Rs$-vectors in the vector subspace $\big\langle H_\rho \big\rangle$, written as
\begin{equation}\label{eq_bj_separte}
  \vec{b}_j \triangleq \begin{bmatrix} \vec{b}^{\,(1)}_j \\ \vec{b}^{\,(2)}_j\\\vdots\\\vec{b}^{\,(s)}_j \end{bmatrix}
  ~~\text{with}~~\vec{b}^{\,(i)}_j \in \Fq^{R},~\forall~1\leq i \leq s,
\end{equation}
such that $\vec{b}^{\,(i)}_1, \vec{b}^{\,(i)}_2, \cdots, \vec{b}^{\,(i)}_{R-r}$ are linearly independent for all $1 \leq i \leq s$, and
\begin{equation}\label{eq_va_HW=0}
  \big\langle \vec{b}_1, \vec{b}_2, \cdots, \vec{b}_{R-r} \big\rangle \cap \big\langle \vec{h}_e:~e \in W\big\rangle = \{\vec{0}\}, ~~\forall~W \in \mW_r.
\end{equation}
For each $1 \leq i \leq s$, we further choose $r$ column $R$-vectors over $\Fq$, denoted by $\vec{b}^{\,(i)}_{R-r+1}, \vec{b}^{\,(i)}_{R-r+2}, \cdots, \vec{b}^{\,(i)}_{R}$, such that $\vec{b}^{\,(i)}_{1}, \vec{b}^{\,(i)}_{2}, \cdots, \vec{b}^{\,(i)}_{R}$ are linearly independent. Now, we let
\begin{equation*}
  B_i \triangleq \Big[ \vec{b}^{\,(i)}_1 ~~\vec{b}^{\,(i)}_2~~\cdots ~~\vec{b}^{\,(i)}_{R}\Big]^{-1}, ~~\forall~1 \leq i \leq s,
\end{equation*}
where the matrix $\Big[ \vec{b}^{\,(i)}_1 ~~\vec{b}^{\,(i)}_2~~\cdots ~~\vec{b}^{\,(i)}_{R}\Big]$ is invertible because its columns are linearly independent by construction. The existence of such $R-r$ vectors $\vec{b}_1,\vec{b}_2, \cdots, \vec{b}_{R-r}$ will be justified in the next subsection (Section~\ref{subsec_exis_b}). In the following, we show that the  matrices $B_i,~1 \leq i \leq s$ thus obtained satisfy \eqref{eq_bR_in_Grho} and~\eqref{eq_bR_cap_G=0}. By recalling the matrix $T$ in~\eqref{eq_T_in_construct}, we have
\begin{align*}
  \widehat{B}^{-1} \cdot T &= \Big[ B^{-1}_1~~B^{-1}_2~~\cdots~~B^{-1}_s  \Big]_s^{\text{diagonal}} \cdot T \\
&=\begin{bmatrix}
B_1^{-1} & \bfzero & \cdots & \bfzero &\\
\bfzero & B_2^{-1} & \cdots & \bfzero &\\
\cdots  & \cdots & \cdots & \cdots\\
\bfzero & \bfzero & \cdots & B_s^{-1}\\
\end{bmatrix} \cdot \begin{bmatrix}
T^{(1)}\\
T^{(2)} \\
\vdots  \\
T^{(s)} \\
\end{bmatrix}\\
&=\begin{bmatrix}
    \vec{b}^{\,(1)}_1 & \vec{b}^{\,(1)}_2 & \cdots & \vec{b}^{\,(1)}_{R-r} \\
    \vec{b}^{\,(2)}_1 & \vec{b}^{\,(2)}_2 & \cdots & \vec{b}^{\,(2)}_{R-r} \\
    \vdots & \vdots & \cdots & \vdots\\
     \vec{b}^{\,(s)}_1 &  \vec{b}^{\,(s)}_2 & \cdots &  \vec{b}^{\,(s)}_{R-r}
  \end{bmatrix} \\
  & = \begin{bmatrix} \vec{b}_1 & \vec{b}_2 & \cdots & \vec{b}_{R-r}\end{bmatrix}.
\end{align*}
Together with the fact that $\vec{b}_1,\,\vec{b}_2, \,\cdots,\, \vec{b}_{R-r}$ are chosen in $\langle H_{\rho}\rangle$ and \eqref{eq_va_HW=0}, we can readily see that the conditions~\eqref{eq_bR_in_Grho} and \eqref{eq_bR_cap_G=0} are satisfied.

To end this subsection, we remark that if the size of the field $\Fq$ of the algebraic sum is too small to be applicable to our code construction, we can take an extension field $\mathbb{F}_{q^L}$ of $\Fq$ such that the size $q^L$ of the extension field is sufficient for constructing an $\mathbb{F}_{q^L}$-valued admissible $(R-r, 1)$ linear secure network code by our code construction. Note that the extension field $\mathbb{F}_{q^L}$ can be viewed as an $L$-dimensional vector space over $\Fq$ with the basis $\{1, \alpha, \alpha^2, \cdots, \alpha^{L-1} \}$, where $\alpha$ is a primitive element of $\mathbb{F}_{q^L}$. Hence, the $\mathbb{F}_{q^L}$-valued $(R-r, 1)$ linear secure network code for computing the algebraic sum on $\mathbb{F}_{q^L}$ can be regarded as an $\big( (R-r)L, L \big)$ linear secure network code for computing the algebraic sum on $\Fq$, which also has the same secure computing rate $R-r$. Explicit upper bounds on the minimum required field size for our code construction will be given in the next subsection. Hence, for the algebraic sum over any finite field, using our code construction, we can always construct an $\Fq$-valued admissible (vector-) linear secure network code of rate up to $C_{\min}-r$ for the model $\langle \mN, f, r \rangle$ with security level $0 \leq r \leq C_{\min}$. This implies the lower bound $C_{\min}-r$ on the secure computing capacity $\hmC\langle \mN, f, r \rangle$, which is formally stated in the following theorem.

\begin{thm}\label{thm_lower_capa}
Consider the model of secure network function computation $\langle \mN, f, r \rangle$, where the target function $f$ is the algebraic sum over a finite field $\Fq$ and the security level $r$ satisfies $0 \leq r \leq C_{\min}$. Then
  \begin{align*}
    \hmC\langle\mN, f, r\rangle \geq C_{\min} - r.
  \end{align*}
\end{thm}

\subsection{The Existence of $\vb_1, \vb_2, \cdots, \vb_{R-r}$ and the Required Field Size of the Code Construction}\label{subsec_exis_b}

We continue to consider the admissible $\Fq$-valued $(R, 1)$ linear network code $\mbC$ for the model $(\mN, f)$, of which the global encoding vectors are $\vec{h}_e \in \Fq^{Rs}$ for all $e \in \mE$. In the following, we will prove that there exist $R-r$ column $Rs$-vectors $\vec{b}_1, \vec{b}_2, \cdots, \vec{b}_{R-r}$ in the vector space $\langle H_\rho\rangle$ such that \rmnum{1}) the condition \eqref{eq_va_HW=0} is satisfied; and \rmnum{2}) $\vec{b}^{\,(i)}_1, \vec{b}^{\,(i)}_2, \cdots, \vec{b}^{\,(i)}_{R-r}$ are linearly independent for each $1 \leq i \leq s$ (cf.~\eqref{eq_bj_separte} for $\vb^{\,(i)}_j$) provided that the field size $q$ satisfies
 \begin{equation*}
   q>|\mW_r|+s.
 \end{equation*}

First, we introduce some notations. Let
\begin{align*}
\mH_W &=\big\langle H_W\big\rangle = \big\langle  \vh_e:~e\in W \big\rangle,\quad \forall~W \in \mW_r\\
\mB_{j} &=\big\langle \vb_1,~\vb_2,~\cdots,~\vb_{j} \big\rangle, ~\quad \qquad \forall~ 1 \leq j \leq R-r \\
\mB^{(i)}_{j} &= \big\langle \vb^{\,(i)}_1,~\vb^{\,(i)}_2,~\cdots,~\vb^{\,(i)}_{j} \big\rangle, ~\,\quad \forall~ 1 \leq j \leq R-r~~\text{and}~~1 \leq i \leq s,
\end{align*}
and let $\mB_{0} = \big\{\vec{0} \in \Fq^{Rs}\big\}$ and $\mB^{(i)}_{0} = \big\{\vec{0} \in \Fq^{R}\big\}$ for all $1 \leq i \leq s$. Further, for $1 \leq i \leq s$ and $0 \leq j \leq R-r$, we define the following set of $\Fq$-valued column $Rs$-vectors:
\begin{align*}
\mathcal{D}^{\langle i \rangle}_j= \left\{\vec{b} \triangleq \left[\begin{smallmatrix}
                                               \\
                                               \vec{b}^{\,(1)} \\
                                               \vec{b}^{\,(2)}\\
                                               \vdots \\
                                               \vec{b}^{\,(s)}\\
                                             \end{smallmatrix}\right]:~~\vec{b}^{\,(i)} \in \mB^{(i)}_{j} ~~\text{and}~~ \vec{b}^{\,(i')} \in \Fq^R~\text{for all}~1 \leq i' \leq s~~\text{with}~~i'\neq i\right\}.
\end{align*}
We can easily verify that $\mathcal{D}^{\langle i \rangle}_j$ is a linear space over $\Fq$.

Now, we choose $R-r$ vectors $\vb_1$, $\vb_2$, $\cdots$, $\vb_{R-r}$ in $\langle H_\rho\rangle$ sequentially, with
  \begin{align}\label{vb_j-1}
\vb_j \in
 \big\langle H_\rho\big\rangle
 \setminus \left[\Big(\bigcup_{W \in \mW_r}(\mH_W + \mB_{j-1}) \Big) \cup  \Big(\bigcup_{i=1}^s \mathcal{D}^{\langle i \rangle}_{j-1} \Big) \right],~~\text{for}~~j=1, 2, \cdots, R-r.
\end{align}
Immediately, we have
\begin{align*}
\vb_j \in \big\langle H_\rho\big\rangle
 \setminus  \bigcup_{W \in \mW_r}(\mH_W + \mB_{j-1}), ~~\forall~1 \leq j \leq R-r,
\end{align*}
which implies \eqref{eq_va_HW=0}. It also follows from~\eqref{vb_j-1} that
\begin{align*}
\vb_j \in \big\langle H_\rho\big\rangle
 \setminus  \bigcup_{i=1}^s \mathcal{D}^{\langle i \rangle}_{j-1}, ~~\forall~1 \leq j \leq R-r.
\end{align*}
This implies that
\begin{align*}
\vb^{\,(i)}_1 \neq \vec{0},~~1\leq i \leq s, \quad\text{and}\quad  \vb_j \notin \bigcup_{i=1}^s \mathcal{D}^{\langle i \rangle}_{j-1},~~2 \leq j \leq R-r,
\end{align*}
and thus for each $1 \leq i \leq s$, the $R-r$ column $R$-vectors $\vec{b}^{\,(i)}_1, \vec{b}^{\,(i)}_2, \cdots, \vec{b}^{\,(i)}_{R-r}$ are linearly independent over $\Fq$.

It now remains to prove that the set on the RHS of~\eqref{vb_j-1} is nonempty for each $1 \leq j \leq R-r$ provided that $q>|\mW_r| + s$. First, by the admissibility of the $(R, 1)$ linear network code $\mbC$ for the model~$(\mN, f)$, i.e., the sink node $\rho$ can compute the algebraic sum $f$ with zero error $R$ times by using the code~$\mbC$ once, there exists a matrix $N$ of size $|\ein(\rho)| \times R$ such that
  \begin{equation*}
  H_\rho \cdot N =     \begin{bmatrix}
      H^{\,(\sigma_1)}_\rho \cdot N \\
      H^{\,(\sigma_2)}_\rho \cdot N \\
      \vdots \\
      H^{\,(\sigma_s)}_\rho \cdot N
    \end{bmatrix} =
    \begin{bmatrix}
      I_R \\
      I_R \\
      \vdots \\
      I_R
    \end{bmatrix}.
  \end{equation*}
  This immediately implies that $\dim \big( \langle H_\rho \big\rangle \big)  = \Rank(H_\rho) \geq R$. Further, for each $W \in \mW_r$ and each $1 \leq j \leq R-r$, we have
  \begin{align}
    &\dim\left(\big(\mH_W + \mB_{j-1}\big) \cap \big\langle H_\rho\big\rangle \right) \nonumber\\
    &\leq \dim\left(\mH_W + \mB_{j-1}\right) \nonumber\\
    &\leq \dim\left(\mH_W \right) + \dim\left(\mB_{j-1}\right) \nonumber\\
    &\leq |W| + j-1 \leq r+j-1\leq R-1.\label{eq_hypothesis}
  \end{align}
  On the other hand, we can see that
  \begin{equation*}
    \dim\left(\big\langle H^{(\,\sigma_i)}_\rho \big\rangle \cap \mB^{(i)}_{j-1}\right) \leq \dim\left(\big\langle \mB^{(i)}_{j-1} \big\rangle\right) \leq R-r-1, \quad \forall~1 \leq i \leq s \text{ and } 1 \leq j \leq R-r.
  \end{equation*}
   This implies that
  \begin{align}
    \dim\left(\big\langle H_\rho \big\rangle \cap \mathcal{D}^{\langle i \rangle}_{j-1}\right) \leq  \dim\left(\big\langle H^{(\,\sigma_i)}_\rho \big\rangle \cap \mB^{(i)}_{j-1}\right) \leq R-r-1,~~\forall~1 \leq i \leq s \text{ and } 1 \leq j \leq R-r.\label{eq_hypothesis2}
  \end{align}
 By the above discussion, for each $1 \leq j \leq R-r$, we have
 \begin{align}
   &\left| \big\langle H_\rho\big\rangle
 \setminus \left[\Big(\bigcup_{W \in \mW_r}(\mH_W + \mB_{j-1}) \Big) \cup  \Big(\bigcup_{i=1}^s \mathcal{D}^{\langle i \rangle}_{j-1} \Big) \right] \right| \nonumber\\
 & \geq   \left|\big\langle H_\rho\big\rangle\right| - \left[\,\sum_{W \in \mW_r}\left|(\mH_W + \mB_{j-1}) \cap \big\langle H_\rho\big\rangle \right| +  \sum_{i=1}^s \left| \mathcal{D}^{\langle i \rangle}_{j-1} \cap \big\langle H_\rho\big\rangle\right|  \right] \nonumber\\
&\geq  q^R - \left[\,\sum_{W \in \mW_r}q^{R-1} +  \sum_{i=1}^s q^{R-r-1}  \right] \label{eq_subspace_reduction} \\
&> q^R - \left[\,\sum_{W \in \mW_r}q^{R-1} +  \sum_{i=1}^s q^{R-1} \right]\nonumber\\
&= q^{R-1} \cdot (q - |\mW_r| - s) \nonumber\\
&>0,\label{eq_use_q>Wr+s}
 \end{align}
  where \eqref{eq_subspace_reduction} follows from \eqref{eq_hypothesis} and \eqref{eq_hypothesis2}, and \eqref{eq_use_q>Wr+s} follows because $q > |\mW_r| + s$. We thus have proved the existence of such $R-r$ vectors $\vec{b}_1, \vec{b}_2, \cdots, \vec{b}_{R-r}$.

In fact, by applying a similar graph-theoretic approach developed in Part~\Rmnum{1} of the current paper \cite{PartI}, we can obtain an improved upper bound on the minimum required field size of our code construction. This improved bound, which is graph-theoretic, depends only on the network topology and the required security level. Before presenting the improved bound, we need to introduce some graph-theoretic notations developed in \cite{PartI}.

In the graph $\mG$, we consider an edge subset $W$ and let $W'$ be a minimum cut separating $W$ from the subset of source nodes $D_W$ (cf.~\eqref{eq_notation_DIJ} for $D_W$). By Lemma~5 in \cite{PartI}, we have $D_W = D_{W'}$. Further, we say that a minimum cut separating $W$ from~$D_W$ is {\em primary} if it separates from $D_W$ all the minimum cuts that separate $W$ from $D_W$, and we use $\widehat{W}$ to denote the primary minimum cut separating~$W$ from~$D_W$. The existence and uniqueness of the primary minimum cut were proved by Guang and Yeung~\cite{GY-SNC-Reduction} (also cf.~\cite{PartI}). Furthermore, we say that an edge subset $W \subseteq \mE$ is {\em primary} if~$W$ is the primary minimum cut separating $W$ itself from $D_W$. We readily see that $W$ is primary if and only if $W=\widehat{W} $. Now, we present the  the improved bound in the theorem below, which shows that with the same target-function-security constraint \eqref{eq_sec_condition_for_f}, it suffices to consider a considerably reduced collection of wiretap sets rather than $\mW_r$.

\begin{thm}\label{thm-enhanced-field-size}
Consider an arbitrary secure network code $\hmbC$ for the model $\langle \mN, f, r \rangle$, where the target function $f$ is the algebraic sum over a finite field $\Fq$ and the security level $r$ satisfies $0 \leq r \leq C_{\min}$. Let
 \begin{align*}
\mW^*_r = \big\{W \in \mW_r:~ W=\widehat{W} \text{ with } |W|=r \big\}.
\end{align*}
 Then, the target-function-security condition \eqref{eq_sec_condition_for_f} is satisfied for the code $\hmbC$ if and only if
 \begin{align*}
        I\big(\vY_W; f(\vM_S)\big) = 0,~~\forall~W \in \mW_r^*.
\end{align*}
\end{thm}

\begin{IEEEproof}
The ``only if'' part is evident because $\mW_r^* \subseteq \mW_r$, and we only need to prove the ``if'' part. We consider an arbitrary wiretap set $W\in \mW_r$. It is not difficult to see that there exists an edge subset $W'$ such that $W'$ separates $W$ from $D_W$ and $|W'|=\mincut(D_{W'},W')=r$, where $\mincut(D_{W'},W')$ is the minimum cut capacity separating $W'$ from $D_{W'}$. We now consider the primary minimum cut $\widehat{W'}$ separating~$W'$ from $D_{W'}$. By the discussion in the paragraph immediately above Theorem~\ref{thm-enhanced-field-size}, we have $D_{\widehat{W'}} = D_{W'}$ (also see~\cite[Lemma 5]{PartI}). Together with the fact that
$$\big|\widehat{W'}\big|=\mincut\big(D_{W'},W'\big)=\big|W'\big|=r,$$
we have $\widehat{W'} \in \mW^*_r$ and thus $I\big(\vY_{\widehat{W'}} ; f(\vM_S)\big)=0$. This further implies that $I\big(\vY_W ; f(\vM_S)\big)=0$ because~$\vY_W$ is a function of $\vY_{\widehat{W'}}$ by the mechanism of network coding. The theorem is thus proved.
\end{IEEEproof}

\medskip

For our code construction developed in this paper, in order to guarantee the condition~\eqref{eq_bR_cap_G=0} for target-function security, by Theorem~\ref{thm-enhanced-field-size}, it is equivalent to guarantee that
\begin{equation*}
    \big\langle \widehat{B}^{-1} \cdot \matT  \big\rangle \cap   \big\langle  H_W \big\rangle = \{\vec{0}\},~~\forall~W \in \mW^*_r
  \end{equation*}
  (cf.~{\bf Verification of the Target-Function-Security Condition} in Section~\ref{subsec_code_construction}). With this, we can reduce the upper bound on the minimum required field size of our code construction from $|\mW_r|+s$ to $|\mW_r^*| + s$. This is formally stated in the following theorem.

  \begin{thm}\label{thm_field_reduc}
Consider the model of secure network function computation $\langle \mN, f, r\rangle$, where the target function $f$ is the algebraic sum over a finite field $\Fq$ and the security level $r$ satisfies $0 \leq r \leq C_{\min}$. Then, for any nonnegative integer $R$ with $r \leq R \leq C_{\min}$, there exists an $\Fq$-valued admissible $(R-r, 1)$ linear secure network code for $\langle \mN, f, r\rangle$ if the field size $q$ satisfies $q> |\mW^*_r| + s$.

\end{thm}

\subsection{An Example}

In this subsection, we will give an example to illustrate our code construction. This example shows that the codes obtained by our code construction, which are target-function secure, are in general not source secure. Also, a field of size smaller than the bound given in Theorem~\ref{thm_field_reduc} is used in our code construction, showing that for our code construction, the bound $|\mW^*_r| + s$ in Theorem~\ref{thm_field_reduc} on the field size is only sufficient but far from being necessary.

\begin{example}\label{eg2}

We consider a secure model $\langle \mN, f, r \rangle$, where $\mN=(\mG, S, \rho)$ is the reverse butterfly network as depicted in Fig.~\ref{fig:butterfly_network2} with the set of source nodes $S=\{\sigma_1,\sigma_2\}$, the target function $f$ being the algebraic sum over the finite field $\mathbb{F}_3$, and the security level $r=1$. By Theorem~\ref{thm_general_upper}, we can obtain that $\hmC\langle \mN, f, r \rangle \leq 1$. Next, we will construct an admissible $(1, 1)$ linear secure network code $\hmbC$ (i.e., $R(\hmbC)=1$) for the model $\langle\mN, f, r\rangle$ by our code construction. This implies that $\hmC\langle\mN, f, r\rangle = 1$ and thus the code $\hmbC$ constructed is optimal.

\begin{figure}[t]
  \centering
{
 \begin{tikzpicture}[x=0.6cm]
    \draw (-3,0) node[vertex] (1) [label=above:{$\sigma_1:\,(x_{1}~x_{2})$}] {};
    \draw ( 3,0) node[vertex] (2) [label=above:{$\sigma_2:\,(y_{1}~y_{2})$}] {};

    \draw ( 0,-1.5) node[vertex] (3) [label=left:] {};
    \draw (-3,-5) node[vertex] (4) [label=left:] {};
    \draw ( 3,-5) node[vertex] (5) [label=right:] {};
    \draw ( 0,-3.5) node[vertex] (6) [label=right:] {};
    \draw ( 0,-6.5) node[vertex] (7) [label=below: {$\rho:\,(x_1+y_1,~x_2+y_2)$}] {};

    \draw[->,>=latex] (1) -- (4) node[midway, auto,swap, pos=0.3, right=-1mm] {$e_1$};
    \draw[->,>=latex] (1) -- (4) node[midway, auto,swap, pos=0.6, left] {$x_1+2x_2$};

    \draw[->,>=latex] (1) -- (3) node[midway, auto, pos=0.2, right=0mm] {$e_2$};
    \draw[->,>=latex] (1) -- (3) node[midway, auto, pos=0.7, left=0.5mm] {$x_2$};

    \draw[->,>=latex] (2) -- (3) node[midway, auto, pos=0.2, left=0mm] {$e_3$};
    \draw[->,>=latex] (2) -- (3) node[midway, auto, pos=0.7, right=0.5mm] {$y_1$};

    \draw[->,>=latex] (2) -- (5) node[midway, auto, pos=0.3, left=-1mm] {$e_4$};
    \draw[->,>=latex] (2) -- (5) node[midway, auto, pos=0.6, right] {$2y_1+y_2$};

    \draw[->,>=latex] (3) -- (6) node[midway, auto, pos=0.3, left=-1mm] {$e_5$};
    \draw[->,>=latex] (3) -- (6) node[midway, auto, pos=0.7, right] {$x_2+y_1$};

    \draw[->,>=latex] (6) -- (4) node[midway, auto, left=-0.5mm] {$e_6$};
     \draw[->,>=latex] (6) -- (4) node[midway, sloped, swap, below] {$x_2+y_1$};

    \draw[->,>=latex] (6) -- (5) node[midway, auto,swap, right=-0.5mm] {$e_7$};
     \draw[->,>=latex] (6) -- (5) node[midway, sloped,swap, below] {$x_2+y_1$};

    \draw[->,>=latex] (4) -- (7) node[midway, auto,swap, right=-0.5mm] {$e_8$};
    \draw[->,>=latex] (4) -- (7) node[midway, sloped,swap, below] {$x_1+y_1$};

    \draw[->,>=latex] (5) -- (7) node[midway, auto, left=-0.5mm] {$e_9$};
     \draw[->,>=latex] (5) -- (7) node[midway, sloped,swap, below] {$x_2+y_2$};
    \end{tikzpicture}
}
\caption{An $\mathbb{F}_3$-valued $(2, 1)$ linear network code for the model $(\mN, f)$.}
  \label{fig:butterfly_network2}
\end{figure}
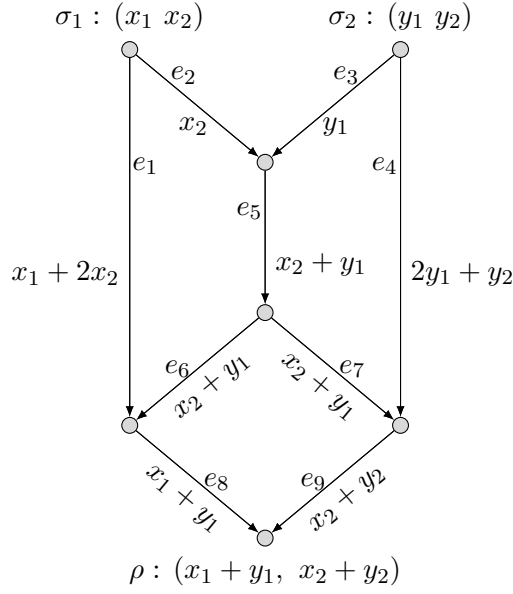

First, we consider an (optimal) $\mathbb{F}_3$-valued $(2,1)$ linear network code $\mbC$ (i.e., $R(\mbC)= 2$) for the model $(\mN, f)$, of which the global encoding vectors are
\begin{align*}
\begin{split}
&\vh_{e_1}=\left[\begin{smallmatrix} 1 \\ 2 \\ 0 \\ 0 \end{smallmatrix}\right], \quad \vh_{e_2} =\left[\begin{smallmatrix} 0\\1 \\ 0 \\ 0 \end{smallmatrix}\right], \quad
\vh_{e_3}=\left[\begin{smallmatrix}  0 \\ 0 \\ 1 \\ 0 \end{smallmatrix}\right], \quad
\vh_{e_4}= \left[\begin{smallmatrix}  0 \\ 0 \\ 2 \\ 1 \end{smallmatrix}\right], \\
&
\vh_{e_5}=\vh_{e_6}=\vh_{e_7}=\left[\begin{smallmatrix}  0 \\ 1 \\ 1 \\ 0 \end{smallmatrix}\right], \quad
\vh_{e_8}=\left[\begin{smallmatrix}  1 \\ 0\\ 1 \\ 0 \end{smallmatrix}\right], \quad
\vh_{e_9}=\left[\begin{smallmatrix}  0 \\ 1 \\ 0 \\ 1 \end{smallmatrix}\right].
\end{split}
\end{align*}
We can readily check that the code $\mbC$ can compute the algebraic sum $f$ twice, but it is not target-function secure because, e.g., $x_1+y_1$ can be determined when the edge $e_8$ is eavesdropped (see Fig.\;\ref{fig:butterfly_network2}).

Based on $\mbC$, we now construct an $\mathbb{F}_3$-valued $(1,1)$ linear secure network code $\hmbC$ for $\langle\mN, f, r\rangle$. Here, $R(\hmbC) =R-r = 2-1=1$. By our code construction, we need to construct a $4 \times 4$ ($Rs=4$) matrix
\begin{equation*}
  \widehat{B} = \Big[B_1~~B_2\Big]_2^{\text{\rm diagonal}} = \begin{bmatrix}
                                                               B_1 & \bfzero \\
                                                               \bfzero & B_2
                                                             \end{bmatrix}
\end{equation*}
with $B_i^{-1}\triangleq \left[\vb_1^{\,(i)}~~\vb_2^{\,(i)}\right]$ for $i=1, 2$, two invertible matrices of size $2 \times 2$, such that the conditions \eqref{eq_bR_in_Grho} and \eqref{eq_bR_cap_G=0} are satisfied. Toward this end, by our code construction it is equivalent to choose
\begin{equation*}
  \vb_1 \triangleq\begin{bmatrix}
                    \vb_1^{\,(1)} \\
                    \vb_1^{\,(2)}
                  \end{bmatrix} ~~\text{and}~~  \vb_2 \triangleq\begin{bmatrix}
                    \vb_2^{\,(1)} \\
                    \vb_2^{\,(2)}
                  \end{bmatrix}
\end{equation*}
sequentially as follows:
\begin{equation*}
  \vb_1 \in \big\langle H_\rho\big\rangle \setminus \left[\bigcup_{e \in \mE}\big\langle\vh_e\big\rangle \cup \mD_0^{\langle 1\rangle} \cup \mD_0^{\langle 2\rangle}\right]
\end{equation*}
(cf.~\eqref{vb_j-1}), namely to choose $\vb_1 \in \big\langle H_\rho\big\rangle = \big\langle \vh_{e_8},\, \vh_{e_9}\big\rangle$ such that $\big\langle \vb_1 \big\rangle \cap \big\langle \vh_e\big\rangle = \{\vec{0}\}$ for all $e \in \mE$ and $\vb_1^{\,(i)} \neq \vec{0}$ for $i=1, 2$; and
$$\vb_2^{\,(i)} \in \Fq^2 \setminus \big\langle \vb_1^{\,(i)}\big\rangle,~~i=1, 2,$$
namely to choose $\vb_2$ such that $\vb_2^{\,(i)}$ and $\vb_1^{\,(i)}$ are linearly independent for $i=1, 2$.
For instance, we can choose
\begin{align*}
\vec{b}_1=\left[\begin{smallmatrix} 1 \\ 2 \\ 1 \\ 2 \end{smallmatrix}\right],~~\text{i.e.},~~\vb_1^{\,(1)}=\vb_1^{\,(2)} = \left[\begin{smallmatrix} 1 \\ \\ 2 \\  \end{smallmatrix}\right],
\end{align*}
and
\begin{align*}
\vec{b}_2^{\,(1)}=\left[\begin{smallmatrix} 0 \\ \\ 1\end{smallmatrix}\right],~~~~\vec{b}_2^{\,(2)}=\left[\begin{smallmatrix}1 \\ \\ 0\end{smallmatrix}\right],~~\text{i.e.,}~~ \vec{b}_2=\left[\begin{smallmatrix} 0 \\ 1 \\1\\0\end{smallmatrix}\right].
\end{align*}
Then, we have
\begin{align*}
 B_1^{-1} = \Big[ \vec{b}_1^{\,(1)} \ \  \vec{b}_2^{\,(1)} \Big]=\begin{bmatrix} 1 & 0  \\ 2 & 1 \end{bmatrix}~~\text{and}~~B_2^{-1} = \Big[ \vec{b}_1^{\,(2)} \ \  \vec{b}_2^{\,(2)} \Big]=\begin{bmatrix} 1 & 1  \\ 2 & 0 \end{bmatrix},
\end{align*}
and thus,
\begin{align*}
 B_1 = \Big[ \vec{b}_1^{\,(1)} \ \  \vec{b}_2^{\,(1)} \Big]^{-1} = \begin{bmatrix} 1 & 0  \\ 1 & 1 \end{bmatrix}~~\text{and}~~B_2 =\Big[ \vec{b}_1^{\,(2)} \ \  \vec{b}_2^{\,(2)} \Big]^{-1} =\begin{bmatrix} 0 & 2  \\ 1 & 1 \end{bmatrix}.
\end{align*}
Also,
\begin{align*}
\widehat{B} = \Big[ B_1~~B_2  \Big]_2^{\text{\rm diagonal}}
=
\begin{bmatrix}
1 & 0&0&0 \\
1 & 1 &0&0\\
0&0&0&2\\
0&0&1&1
\end{bmatrix}.
\end{align*}
According to the code construction, we now obtain an $\mathbb{F}_3$-valued $(1,1)$ linear secure network code $\hmbC$ for $\langle \mN, f, r \rangle$, where the global encoding vectors are $\vg_{e}=\widehat{B}  \cdot \vh_e$ for all $e\in \mE$, i.e.,
\begin{align}\label{sec_global_kernels_EX}
\begin{split}
&\vg_{e_1}=\left[\begin{smallmatrix} 1 \\ 0\\ 0 \\ 0 \end{smallmatrix}\right], \quad \vg_{e_2} =\left[\begin{smallmatrix} 0\\1 \\ 0 \\ 0 \end{smallmatrix}\right], \quad
\vg_{e_3}=\left[\begin{smallmatrix}  0 \\ 0 \\ 0 \\ 1 \end{smallmatrix}\right], \quad
\vg_{e_4}=\left[\begin{smallmatrix}  0 \\ 0 \\ 2\\ 0 \end{smallmatrix}\right], \\
&
\vg_{e_5}=\vg_{e_6}=\vg_{e_7}=\left[\begin{smallmatrix}  0 \\ 1 \\ 0 \\ 1 \end{smallmatrix}\right], \quad
\vg_{e_8}=\left[\begin{smallmatrix}  1 \\ 1\\ 0 \\ 1 \end{smallmatrix}\right], \quad
\vg_{e_9}=\left[\begin{smallmatrix}  0 \\ 1 \\ 2 \\ 1 \end{smallmatrix}\right].
\end{split}
\end{align}

In using the code $\hmbC$, let $m_i$ and~$k_i$ in $\mathbb{F}_3$ be the source message and key generated by the source node~$\sigma_i$, $i=1,2$. We use $y_e$, taking values in $\mathbb{F}_3$, to denote the message transmitted on each edge $e\in\mE$. By the global encoding vectors of $\hmbC$ in \eqref{sec_global_kernels_EX}, the messages $y_e$ transmitted on the edges $e\in\mE$ are
\begin{align}\label{equ-algebraic_sum_keys}
\begin{split}
&y_{e_1}=m_1, \quad y_{e_2}=k_1,\quad y_{e_3}=k_2, \quad y_{e_4}=2m_2,\\
&y_{e_5}=y_{e_6}=y_{e_7}=k_1+k_2, \quad y_{e_8}=m_1+k_1+k_2,\quad y_{e_9}=2m_2+k_1+k_2.
\end{split}
\end{align}
We can readily verify the computability and target-function-security conditions for the code $\hmbC$. To be specific, by using the two messages $y_{e_8}$ and $y_{e_9}$ received at the sink node $\rho$, the algebraic sum $m_1+m_2$ over $\mathbb{F}_3$ can be computed with zero error by calculating $y_{e_8} + 2y_{e_9}$. On the other hand, it is easy to check that the wiretapper cannot obtain any information about $m_1+m_2$ when any one edge is eavesdropped.

Furthermore, we mention that the code $\hmbC$ is not source secure because, e.g., $m_1$  is obtained when the edge $e_1$ is eavesdropped.
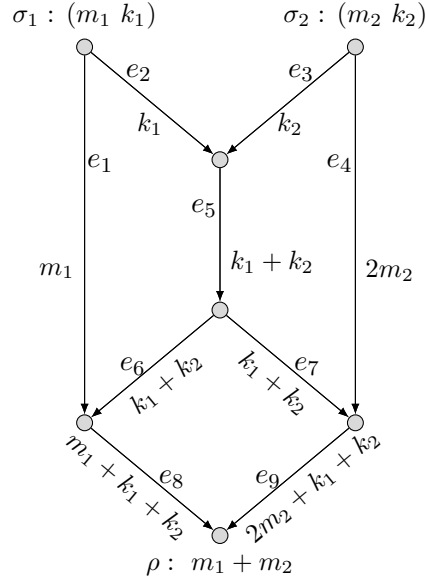
\begin{figure}[t]
\centering
 \begin{tikzpicture}[x=0.6cm]
    \draw (-3,0) node[vertex] (1) [label=above:{\small $\sigma_1:\,(m_{1}~k_{1})$}] {};
    \draw ( 3,0) node[vertex] (2) [label=above:{\small $\sigma_2:\,(m_{2}~k_{2})$}] {};

    \draw ( 0,-1.5) node[vertex] (3) [label=left:] {};
    \draw (-3,-5) node[vertex] (4) [label=left:] {};
    \draw ( 3,-5) node[vertex] (5) [label=right:] {};
    \draw ( 0,-3.5) node[vertex] (6) [label=right:] {};
    \draw ( 0,-6.5) node[vertex] (7) [label=below:\small $\rho:~m_1+m_2$] {};

    \draw[->,>=latex] (1) -- (4) node[midway, auto,swap, pos=0.3, right=-1mm] { $e_1$};
    \draw[->,>=latex] (1) -- (4) node[midway, auto,swap, pos=0.6, left] {\small $m_1$};

    \draw[->,>=latex] (1) -- (3) node[midway, auto, pos=0.2, right=0mm] {$e_2$};
    \draw[->,>=latex] (1) -- (3) node[midway, auto, pos=0.7, left=0.5mm] {\small $k_1$};

    \draw[->,>=latex] (2) -- (3) node[midway, pos=0.2, left=0mm] {$e_3$};
    \draw[->,>=latex] (2) -- (3) node[midway, pos=0.7, right=0.5mm] {\small $k_2$};

    \draw[->,>=latex] (2) -- (5) node[midway, auto, pos = 0.3, left=-1mm] {$e_4$};
    \draw[->,>=latex] (2) -- (5) node[midway, auto, pos=0.6,right] {\small $2m_2$};

    \draw[->,>=latex] (3) -- (6) node[midway, auto, pos=0.3, left=-1mm] {$e_5$};
    \draw[->,>=latex] (3) -- (6) node[midway, auto, pos=0.7, right] {\small $k_1+k_2$};

    \draw[->,>=latex] (6) -- (4) node[midway, auto, left=-0.5mm] {$e_6$};
     \draw[->,>=latex] (6) -- (4) node[midway, sloped, swap, below] {\small $k_1+k_2$};

    \draw[->,>=latex] (6) -- (5) node[midway, auto,swap, right=-0.5mm] {$e_7$};
     \draw[->,>=latex] (6) -- (5) node[midway, sloped,swap, below] {\small $k_1+k_2$};

    \draw[->,>=latex] (4) -- (7) node[midway, auto,swap, right=-0.5mm] {$e_8$};
    \draw[->,>=latex] (4) -- (7) node[midway, sloped, swap, left=2mm, below] {\small $m_1+k_1+k_2$};

    \draw[->,>=latex] (5) -- (7) node[midway, auto, left=-0.5mm] {$e_9$};
     \draw[->,>=latex] (5) -- (7) node[midway, sloped, swap, right=2mm, below] {\small $2m_2+k_1+k_2$};
    \end{tikzpicture}
\caption{An $\mathbb{F}_3$-valued $(1,1)$ linear secure network code for the secure model $\langle \mN, f, r\rangle$.}
  \label{butfly_net_rate_3_over_22}
\end{figure}
\end{example}

For the secure model $\langle\mN, f, r\rangle$ as discussed in the above example, by Theorem~\ref{thm_field_reduc}, we can construct an $\Fq$-valued admissible $(1,1)$ linear secure network code for $\langle\mN, f, r \rangle$ if the field size $q > |\mW_1^*|+  2=9$, where $\mW_1^*=\big\{\{e_1\},\{e_2\},\{e_3\},\{e_4\},\{e_5\},\{e_8\},\{e_9\}\big\}$. However, we see in the example that the finite field $\mathbb{F}_3$ is sufficient for our code construction. This implies that for our code construction, the bound $|\mW_r^*| + s$ in Theorem~\ref{thm_field_reduc} on the field size is sufficient but far from being necessary.

\section{Comparison of the Code Constructions for the Models $\langle \mN, f, r \rangle$ and $(\mN, f, r)$}\label{sec:comparison}

We recall the source-security model $(\mN, f, r)$ investigated in Part~\Rmnum{1} of the current paper \cite{PartI} (also mentioned in Section~\ref{subsec_comp_bound}), where we have developed a code construction for $(\mN, f, r)$ for security level $0 \leq r \leq C_{\min}$. By this code construction we can construct an admissible scalar linear secure network code for $(\mN, f, r)$ with secure computing rate up to $C_{\min}-r$. In this section, we first present a generalized code construction for the source-security model $(\mN, f, r)$, by which more codes can be obtained than that by the previous code construction in \cite{PartI}. Furthermore, we note that an admissible secure network code for $(\mN, f, r)$ is also admissible for $\langle \mN, f, r \rangle$. In particular, the codes constructed by the generalized code construction for $(\mN, f, r)$ are still admissible for $\langle \mN, f, r\rangle$. We further show that all codes constructed by the generalized code construction are only a very special subclass of the codes obtained by our code construction for $\langle \mN, f, r \rangle$, and the ratio of the number of codes that can be constructed by the generalized code construction for $(\mN, f, r)$ to the number of codes that can be constructed by the code construction for $\langle \mN, f, r \rangle$ tends to $0$ as the field size $q$ tends to infinity. Moreover, a smaller field size can be used for our code construction for $\langle \mN, f, r \rangle$ compared with the generalized code construction for $(\mN, f, r)$.

\subsection{A Generalized Code Construction for the Source-Security Model $(\mN, f, r)$}

We continue to consider the source-security model $(\mN, f, r)$ and let $\hmbC$ be an $\Fq$-valued admissible $(\ell, 1)$ linear secure network code for $(\mN, f, r)$, of which the global encoding vector $\vec{g}_e$ for each $e \in \mE$ is of dimension $s\ell + \sum_{i=1}^{s}z_i$. Here, the nonnegative integer $z_i$ for $1\leq i \leq s$ is the dimension of the random key $\vK_i$ available at the source node $\sigma_i$, i.e., $\vK_i$ is uniformly distributed on the vector space $\Fq^{z_i}$. Now, we define a matrix $\Gamma(\hmbC)$ for the code $\hmbC$ by
\begin{align}
    \Gamma(\hmbC) &\triangleq  \Big[ T^{(1)}~~T^{(2)}~~\cdots~~ T^{(s)}\Big]_s^{\text{diagonal}} =  \begin{bmatrix}
    T^{(1)} & \mathbf{0} & \cdots & \mathbf{0} \\
    \mathbf{0} &  T^{(2)}& \cdots & \mathbf{0} \\
    \cdots & \cdots & \cdots & \cdots \\
    \mathbf{0} & \mathbf{0} & \cdots &  T^{(s)}
  \end{bmatrix}\label{eq_GammaCs_def_s},
  \end{align}
where we recall by \eqref{eq_GammaC_def} that $T^{(i)} =\begin{bmatrix}I_\ell \\ \bfzero_{z_i \times \ell} \end{bmatrix}$ for $1 \leq i \leq s$. Then, the matrix $\Gamma(\hmbC)$ is of size $\big(\sum_{i=1}^{s}(\ell+z_i)\big) \times s\ell$. In the following, we write $\Gamma(\hmbC)$ as $\Gamma$ for notational simplicity if there is no ambiguity on the code $\hmbC$. With this, we have
\begin{equation}\label{eq_XSGA=fMS_s}
   \big(\vM_S~~\vK_S\big) \cdot\Gamma =\vM_S,
\end{equation}
where we use $(\vM_S~\vK_S)$ to represent $\big((\vM_1~\vK_1)~(\vM_2~\vK_2)~\cdots~(\vM_s~\vK_s)\big)$.
Then the source-security condition can be written as
\begin{align}\label{eq_equivalent_source_security_condition_s}
        I\big(\vY_W; \vM_S) = I\big(\vY_W;\big(\vM_S~\vK_S\big) \cdot\Gamma\big) = 0,~~\forall~W \in \mW_r.
\end{align}

Motivated by the equivalent condition \eqref{def_sec_condition_space} of the target-function-security condition \eqref{eq_sec_condition_for_f} as presented in Theorem~\ref{thm_sec_condition_space_general}, we can also prove an equivalent condition of the source-security condition \eqref{eq_equivalent_source_security_condition_s} as stated in the following theorem, whose proof is deferred to Appendix~\ref{app_proof_of_thm9}.

\begin{thm}\label{thm_sec_condition_space_general_for_s}
Consider an $(\ell, 1)$ linear secure network code $\hmbC$ for the source-security model $(\mN, f, r)$, where the global encoding vectors are $\vec{g}_e,~e \in\mE$. Then, the source-security condition \eqref{eq_equivalent_source_security_condition_s} is satisfied for the code $\hmbC$ if and only if
  \begin{align}\label{def_sec_condition_space_s}
    \big\langle G_W  \big\rangle \cap \big\langle \Gamma \big\rangle = \big\{\vec{0}\big\},~~\forall~W \in \mW_r.
  \end{align}
\end{thm}

For the source-security model $(\mN, f, r)$ with security level $0 \leq r \leq C_{\min}$, we have developed in~\cite{PartI} a code construction for $(\mN, f, r)$, by which we can construct an admissible $(R-r, 1)$ linear secure network code for $(\mN, f, r)$, where $R$ can take any integer value between $r$ and $C_{\min}$. With Theorem~\ref{thm_sec_condition_space_general_for_s}, we are able to not only generalize this code construction but also present it in a more succinct way.

We continue to consider the source-security model $(\mN, f, r)$ with $0 \leq r \leq C_{\min}$. First, we construct an admissible $(R, 1)$ linear network code $\mbC$ for the model $(\mN, f)$, of which the global encoding vectors are $\vh_e \in \Fq^{Rs}$ for all $e \in \mE$. Here, $R$ can take any integer value betweeen $r$ and $C_{\min}$. Next, we choose an $\Fq$-valued $R \times R$ invertible matrix $A$ and let
\begin{align}\label{eq_Ahat}
\widehat{A} = \Big[ A~~A~~\cdots~~A  \Big]_s^{\text{diagonal}} =\begin{bmatrix}
                                                                  A & \bfzero & \cdots & \bfzero \\
                                                                  \bfzero & A &\cdots & \bfzero \\
                                                                  \vdots & \vdots & \vdots & \vdots \\
                                                                  \bfzero& \bfzero & \cdots & A
                                                                \end{bmatrix}_{Rs \times Rs},
\end{align}
an $s \times s$ invertible block matrix, such that the two conditions
\begin{equation}\label{eq_bR_in_Grho_A}
    \big\langle \widehat{A}^{-1} \cdot \matT  \big\rangle \subseteq \big\langle H_\rho \big\rangle
\end{equation}
and
\begin{equation}\label{eq_bR_cap_G=0_A}
    \big\langle \widehat{A}^{-1} \cdot \Gamma  \big\rangle \cap \big\langle  H_W \big\rangle = \{\vec{0}\},~~\forall~W \in \mW_r
\end{equation}
are satisfied, where we recall \eqref{eq_T_in_construct} for $T$ and \eqref{eq_GammaCs_def_s} for $\Gamma$ with $\ell = R-r$ and $z_i = r$ for $1 \leq i\leq s$. Similar to \eqref{eq_bR_in_Grho} and \eqref{eq_bR_cap_G=0}, the equations \eqref{eq_bR_in_Grho_A} and \eqref{eq_bR_cap_G=0_A} are associated with the computability condition and source-security condition, respectively. With the above, we will prove that the transformation of the code $\mbC$ by the matrix $\widehat{A}$, i.e, $\widehat{A} \cdot \mbC$, is an admissible $(R-r, 1)$ linear secure network code for the source-security model $(\mN, f, r)$.

First, we note that the matrix $\widehat{A}$ in \eqref{eq_bR_in_Grho_A} satisfies all the constraints on $\widehat{B}$ in \eqref{eq_bR_in_Grho}. Then \eqref{eq_bR_in_Grho_A} becomes \eqref{eq_bR_in_Grho} if we regard $\widehat{A}$ as $\widehat{B}$. We thus have verified the computability condition for the code $\widehat{A} \cdot \mbC$ (cf.~\textbf{Verification of the Computability Condition} in Section~\ref{subsec_code_construction}). On the other hand, we note that the equation \eqref{eq_bR_cap_G=0_A} holds if and only if the equation \eqref{def_sec_condition_space_s} holds, i.e.,
\begin{equation*}
   \big\langle \widehat{A}^{-1} \cdot \Gamma  \big\rangle \cap \big\langle  H_W \big\rangle = \{\vec{0}\}~~\text{if and only if}~~\big\langle \Gamma \big\rangle \cap \big\langle G_W  \big\rangle= \big\{\vec{0}\big\},~~\forall~W \in \mW_r,
\end{equation*}
where
\begin{equation*}
  G_W = \Big[\vg_e:~e\in W\Big] = \Big[\widehat{A} \cdot \vh_e:~e \in W\Big] = \widehat{A} \cdot H_W
\end{equation*}
with $\vg_e$ denoting the global encoding vector of $\widehat{A} \cdot \mbC$ for an edge $e \in \mE$. By Theorem~\ref{thm_sec_condition_space_general_for_s}, we have also verified the source-security condition~\eqref{eq_equivalent_source_security_condition_s} for the code $\widehat{A} \cdot \mbC$.

\medskip

Now, we recall the previous code construction developed in \cite{PartI} for $(\mN, f, r)$. Similar to the above construction, we first construct an admissible $(R, 1)$ linear network code $\mbC$ for $(\mN, f)$, of which the global encoding vectors are $\vh_e$ for all $e \in \mE$ (cf.~\eqref{eq_he}). Next, we sequentially choose $R$ linearly independent column $R$-vectors $\va_1, \va_2, \cdots, \va_R$ as follows:
\begin{itemize}
  \item For $1 \leq j \leq R-r$, choose
\begin{align}\label{eq_aj1}
\va_j \in
\Fq^R \setminus \bigcup_{W \in \mW_r}~\bigcup_{i=1}^s \Big(\mH_W^{(i)} + \big\langle \va_1, \va_2, \cdots, \va_{j-1} \big\rangle  \Big),
\end{align}
where $\mH_W^{(i)} \triangleq \big\langle  \vh_e^{\,(\sigma_i)}:~e\in W \big\rangle$ for $1 \leq i \leq s$ and $W\in \mW_r$;
  \item For $R-r+1 \leq j \leq R$, choose
\begin{align}\label{eq_aj2}
\va_j \in
\Fq^R \setminus \big\langle \va_1, \va_2, \cdots, \va_{j-1} \big\rangle.
\end{align}
\end{itemize}
Let $A = \Big[\va_1~~\va_2~~\cdots~~\va_R\Big]^{-1}$ and $\widehat{A} = \Big[ A~~A~~\cdots~~A  \Big]_s^{\text{diagonal}}$ (cf.~\eqref{eq_Ahat}). In Part~\Rmnum{1} of the current paper~\cite{PartI}, we have proved that $\widehat{A} \cdot \mbC$ is an admissible $(R-r, 1)$ linear secure network code for the source-security model $(\mN, f, r)$. We refer the readers to Section~\Rmnum{4}-C therein for more details.

 Next, we claim that choosing the $R$-vectors $\va_1, \va_2, \cdots, \va_R$ according to \eqref{eq_aj1} and \eqref{eq_aj2} is equivalent to choosing an $R \times R$ invertible matrix $A^{-1} = \Big[\va_1~~\va_2~~\cdots~~\va_R\Big]$ such that
\begin{equation}\label{eq_equi_A1-1}
  \big\langle\widehat{A}^{-1}\cdot T\big\rangle \subseteq \big\langle H_{\rho}\big\rangle
\end{equation}
and
\begin{equation}\label{eq_equi_A2}
   \big\langle A^{-1} \cdot T^{(i)} \big\rangle \cap \mH_W^{(i)} = \{\vec{0}\},~~\forall~W \in \mW_r~~\text{and}~~1\leq i \leq s;
\end{equation}
or equivalently,
\begin{equation}\label{eq_equi_A12}
  \left\langle  {\small\begin{bmatrix}A^{-1}_{R-r} \\A^{-1}_{R-r} \\\vdots \\A^{-1}_{R-r}\end{bmatrix}} \right\rangle \subseteq \big\langle H_{\rho}\big\rangle,
\end{equation}
and
\begin{equation}\label{eq_equi_A22}
   \big\langle A^{-1}_{R-r} \big\rangle \cap \mH_W^{(i)} = \{\vec{0}\},~~\forall~W \in \mW_r~~\text{and}~~1\leq i \leq s,
\end{equation}
where $A^{-1}_{R-r} = \Big[\va_1~~\va_2~~\cdots~~\va_{R-r}\Big]$. With this claim, we can readily see that the code construction developed in \cite{PartI} is a special case of our generalized code construction because \eqref{eq_equi_A2} implies \eqref{eq_bR_cap_G=0_A}, but not vice versa.

To prove this claim, we first consider a matrix $A^{-1} = \Big[\va_1~~\va_2~~\cdots~~\va_R\Big]$ obtained by \eqref{eq_aj1} and~\eqref{eq_aj2}, and further let $\widehat{A} = \Big[ A~~A~~\cdots~~A  \Big]_s^{\text{diagonal}}$. We now show that the matrices $A$ and $\widehat{A}$ satisfy \eqref{eq_equi_A12} and~\eqref{eq_equi_A22}. First, by the computability of the $(R, 1)$ linear network code $\mbC$ for $(\mN, f)$, we have
\begin{equation*}
  \left\langle  {\small \begin{bmatrix}I_R\\I_R \\\vdots \\I_R\end{bmatrix}} \right\rangle \subseteq \big\langle H_{\rho}\big\rangle.
\end{equation*}
This immediately implies that
\begin{equation*}
  \left\langle {\small \begin{bmatrix}A^{-1}_{R-r} \\A^{-1}_{R-r} \\\vdots \\A^{-1}_{R-r}\end{bmatrix}} \right\rangle \subseteq \Bigg\langle {\small \begin{bmatrix}I_R\\I_R \\\vdots \\I_R\end{bmatrix}}\Bigg\rangle \subseteq \big\langle H_{\rho}\big\rangle.
\end{equation*}
Thus, \eqref{eq_equi_A12} is always satisfied for any $R \times R$ matrix $A$ so constructed. It further follows from \eqref{eq_aj1} that \eqref{eq_equi_A22} is satisfied.

Next, we consider an $R \times R$ invertible matrix $A^{-1} \triangleq \Big[\va_1~~\va_2~~\cdots~~\va_R\Big]$ that satisfies \eqref{eq_equi_A12} and \eqref{eq_equi_A22}. By \eqref{eq_equi_A22} we obtain that for each $j=1, 2, \cdots, R-r$,
\begin{equation*}
  \big\langle \va_1,~\va_2, \cdots, \va_j \big\rangle \cap \mH_W^{(i)} = \{\vec{0}\},~~\forall~W \in \mW_r~~\text{and}~~1 \leq i \leq s.
\end{equation*}
Thus this implies \eqref{eq_aj1} for $1 \leq j \leq R-r$. Together with the invertibility of $A^{-1}$, \eqref{eq_aj2} is satisfied for $R-r+1 \leq j \leq R$. This proves the claim that choosing the $R$-vectors $\va_1, \va_2, \cdots, \va_R$ according to \eqref{eq_aj1} and \eqref{eq_aj2} is equivalent to choosing an $R \times R$ invertible matrix $A^{-1} = \Big[\va_1~~\va_2~~\cdots~~\va_R\Big]$ that satisfies \eqref{eq_equi_A1-1} and \eqref{eq_equi_A2}.

\subsection{Comparison of Our Code Constructions for $(\mN, f, r)$ and $\langle \mN, f, r  \rangle$}

In this subsection, we compare the generalized code construction for the source-security model $(\mN, f, r)$ and the code construction for the target-function-security model $\langle \mN, f,r \rangle$. We first show that the codes constructed by our generalized code construction for $(\mN, f, r)$ can be also constructed by our code construction for $\langle \mN, f, r \rangle$. To show this, it suffices to prove that an $R \times R$ invertible matrix $A$ that satisfies \eqref{eq_bR_in_Grho_A} and \eqref{eq_bR_cap_G=0_A} also satisfies the conditions \eqref{eq_bR_in_Grho} and \eqref{eq_bR_cap_G=0}, or more precisely,
\begin{equation}\label{eq_bR_in_Grho_A*}
    \big\langle \widehat{A}^{-1} \cdot \matT  \big\rangle \subseteq \big\langle H_\rho \big\rangle,
\end{equation}
and
\begin{equation}\label{eq_bR_cap_G=0_A**}
    \big\langle \widehat{A}^{-1} \cdot \matT  \big\rangle \cap   \big\langle  H_W \big\rangle = \{\vec{0}\},~~\forall~W \in \mW_r.
  \end{equation}
We note that \eqref{eq_bR_in_Grho_A*} is identical to \eqref{eq_bR_in_Grho_A} which has already been proved, so we only need to prove \eqref{eq_bR_cap_G=0_A**}. Assume the contrary that for a wiretap set $W \in \mW_r$,
\begin{equation*}
  \big\langle \widehat{A}^{-1} \cdot T\big\rangle \cap \big\langle H_W \big\rangle  \neq \{\vec{0}\},
\end{equation*}
or equivalently, there exist two non-zero column vectors $\vec{\alpha} \in \Fq^{R-r}$ and $\vec{\beta} \in \Fq^{|W|}$ such that
\begin{equation}\label{eq_app_AinTal = Hwbeta}
  \widehat{A}^{-1} \cdot T \cdot \vec{\alpha} = H_W \cdot \vec{\beta} \neq \vec{0}.
\end{equation}
Next, we consider
\begin{align}
  &\widehat{A}^{-1} \cdot \Gamma \cdot \begin{bmatrix}
                                        \vec{\alpha} \\
                                        \vec{\alpha} \\
                                        \vdots \\
                                        \vec{\alpha}
                                      \end{bmatrix}=\widehat{A}^{-1} \cdot \Big[ T^{(1)}~~T^{(2)}~~\cdots~~T^{(s)}  \Big]_s^{\text{diagonal}} \cdot \begin{bmatrix}
                                        \vec{\alpha} \\
                                        \vec{\alpha} \\
                                        \vdots \\
                                        \vec{\alpha}
                                      \end{bmatrix}\nonumber\\
  &= \widehat{A}^{-1} \cdot\begin{bmatrix}
                                        T^{(1)} \cdot \vec{\alpha} \\
                                        T^{(2)} \cdot \vec{\alpha} \\
                                        \vdots \\
                                        T^{(s)} \cdot \vec{\alpha}
                                      \end{bmatrix}\nonumber\\
  &= \widehat{A}^{-1} \cdot T \cdot \vec{\alpha} \nonumber\\
  &= H_W \cdot \vec{\beta}\nonumber\\
  & \neq \vec{0}\label{eq_app_AinTal = Hwbeta_cont},
\end{align}
where \eqref{eq_app_AinTal = Hwbeta_cont} immediately follows from \eqref{eq_app_AinTal = Hwbeta}. This implies that $\big\langle \widehat{A}^{-1} \cdot \Gamma\big\rangle \cap \big\langle H_W \big\rangle \neq \{\vec{0}\}$, a contradiction to the condition \eqref{eq_bR_cap_G=0_A}.

Furthermore, the $(R-r, 1)$ codes thus constructed for $(\mN, f, r)$, which are also admissible for $\langle \mN, f, r \rangle$, are only a very small subclass of the $(R-r, 1)$ codes constructed by our code construction for $\langle \mN, f, r \rangle$. To show this, we consider the ratio of the number of the $(R-r, 1)$ codes constructed by the generalized code construction for $(\mN, f, r)$ to the number of the $(R-r, 1)$ codes constructed by the code construction for $\langle \mN, f, r \rangle$. In the following we will prove that this ratio tends to 0 as the field size $q$ tends to infinity. Let $\mbC$ be an $\Fq$-valued admissible $(R, 1)$ linear network code for $(\mN, f)$ and let
\begin{align}
  &\widehat{\mathcal{B}} \triangleq \Bigg\{\widehat{B} =\Big[ B_1~B_2~\cdots~B_s  \Big]_s^{\text{diagonal}} :~\text{$\Fq$-valued $R \times R$ invertible matrices}~B_i,~1\leq i \leq s,\nonumber\\
   &\hspace{20mm}\text{s.t.}~\widehat{B} ~\text{satisfies the conditions \eqref{eq_bR_in_Grho} and \eqref{eq_bR_cap_G=0}}\Bigg\},\label{eq_Bhat_set}\\
  &\widehat{\mathcal{A}} \triangleq \Bigg\{\widehat{A} =\Big[ A~A~\cdots~A  \Big]_s^{\text{diagonal}} :~\text{an $\Fq$-valued $R \times R$ invertible matrix}~A \nonumber\\
   &\hspace{20mm}\text{s.t.}~\widehat{A} ~\text{satisfies the conditions  \eqref{eq_bR_in_Grho_A} and \eqref{eq_bR_cap_G=0_A}}\Bigg\}.\nonumber
\end{align}
By the discussions in the last subsection, we know that all the matrices $\widehat{A}$ in $\widehat{\mA}$ and all the constructed $\Fq$-valued admissible $(R-r, 1)$ linear secure network codes $\hmbC$ for $(\mN, f, r)$ building on $\mbC$ are one-to-one corresponding. Similarly, all the matrices $\widehat{B}$ in $\widehat{\mB}$ and all the constructed $\Fq$-valued admissible $(R-r, 1)$ linear secure network codes $\hmbC$ for $\langle\mN, f, r\rangle$ building on $\mbC$  are also one-to-one corresponding.\footnote{See Appendix~\ref{app_2} for the justification of this one-to-one correspondence.}

We first lower bound the size of $\widehat{\mB}$. We use $n_j,~1 \leq j \leq R$ to denote the number of all the possible vectors of $\vec{b}_j$, where
\begin{equation*}
  \vec{b}_j = \begin{bmatrix} \vec{b}^{\,(1)}_j \\ \vec{b}^{\,(2)}_j\\\vdots\\\vec{b}^{\,(s)}_j \end{bmatrix}
\end{equation*}
(cf.~\eqref{eq_bj_separte}). For $1 \leq j \leq R-r$, it follows from \eqref{vb_j-1} and \eqref{eq_use_q>Wr+s} that
\begin{equation*}
  n_j \geq q^{R-1}\cdot(q - |\mW_r| - s).
\end{equation*}
For $R-r+1 \leq j \leq R$, choose $\vb_j^{\,(i)}$ for each $1 \leq i \leq s$ such that $\vb_j^{\,(i)}$ and $\vb_1^{\,(i)},~\vb_2^{\,(i)},~\cdots,~\vb_{j-1}^{\,(i)}$ are linearly independent, or equivalently, choose $\vb_j^{\,(i)}$ such that
\begin{equation*}
  \vb_j^{\,(i)} \in \Fq^R \setminus \big\langle \vb_1^{\,(i)}, ~\vb_2^{\,(i)}, ~\cdots, ~\vb_{j-1}^{\,(i)}\big\rangle,\quad \forall~1\leq i \leq s.
\end{equation*}
This implies that
\begin{equation*}
  n_j = \big(q^R - q^{j-1}\big)^s, ~~R-r+1 \leq j \leq R.
\end{equation*}
Combining the above, we can lower bound the size of $\widehat{\mB}$ by considering
\begin{align}\label{B_lowerbound}
  \big|\widehat{\mB}\big|  = \prod_{i=1}^{R}n_j \geq \prod_{j=1}^{R-r}q^{R-1}(q-|\mW_r|-s) \cdot  \prod_{j=R-r+1}^{R}(q^R - q^{j-1})^s.
\end{align}
Furthermore, we can upper bound the size of $\widehat{\mA}$ trivially by considering
\begin{align}\label{A_upperbound}
  \big|\widehat{\mA}\big| &\leq \prod_{j=1}^{R}q^R = q^{R^2}.
\end{align}
By \eqref{B_lowerbound} and \eqref{A_upperbound}, we can obtain the upper bound
\begin{align*}
  \frac{ \big|\widehat{\mA}\big|}{\big|\widehat{\mB}\big|} &\leq \frac{q^{R^2}}{\prod\limits_{j=1}^{R-r}q^{R-1}(q-|\mW_r|-s)\cdot  \prod\limits_{j=R-r+1}^{R}(q^R - q^{j-1})^s},
\end{align*}
which tends to $0$ as the field size $q$ tends to infinity provided that $s>1$.

\medskip

In addition, we remark that a smaller field size can be used for our code construction for $\langle \mN, f, r \rangle$ compared with our generalized code construction for $(\mN, f, r)$. To be specific, by Theorem~13 in \cite{PartI}, we know that if the field size $q>s \cdot |\mW_r^*|$, an $\Fq$-valued admissible $(R-r, 1)$ linear secure network code for $(\mN, f, r)$ can be constructed by our code construction. By Theorem~\ref{thm_field_reduc}, an $\Fq$-valued admissible $(R-r, 1)$ linear secure network code for $\langle \mN, f, r \rangle$ can be constructed by our code construction if the field size $q >|\mW_r^*| + s$. It is evident that in general $|\mW_r^*| + s$ is considerably smaller than $s \cdot |\mW_r^*|$.

\section{Conclusion}\label{sec:concl}

Secure network function computation has been put forward in this two-part paper. In the current paper (i.e., Part~\Rmnum{2} of the two-part paper), we have investigated securely computing a linear function over a finite field with a wiretapper who can eavesdrop any edge subset up to a certain size but is not allowed to obtain any information about the target function of the linear function. We have proved an upper bound on the secure computing capacity for target-function security, which is applicable to arbitrary network topologies and arbitrary security levels. In particular, when no security is considered, the upper bound reduces to the capacity for computing a linear function on a network without security consideration. Nevertheless, the secure computing capacity may still coincide with the computing capacity even for a non-trivial security level $r>0$, i.e., there is no penalty on the secure computing capacity compared with the computing capacity without security consideration. We further proved that our upper bound on the secure computing capacity for target-function security is always not less than the upper bound obtained in Part~\Rmnum{1} \cite{PartI} for source security, which accords with the relation between the secure computing capacities for target-function security and source security. Next, from an algebraic point of view, we have obtained two equivalent conditions for target-function security and source security for the existence of the corresponding linear secure network codes. Based on the two equivalent conditions, we developed a code construction of linear secure network codes for target-function security and generalized the code construction developed in Part~\Rmnum{1} \cite{PartI} to construct linear secure network codes for source security. Furthermore, we compared the two code constructions and showed that the ratio of the number of codes that can be constructed by the generalized code construction for source security to the number of codes that can be constructed by the code construction for target-function security tends to $0$ as the field size tends to infinity.

\numberwithin{thm}{section}
\appendices

\section{Proof of Theorem~\ref{thm_sec_condition_space_general_for_s}}\label{app_proof_of_thm9}

We can prove Theorem~\ref{thm_sec_condition_space_general_for_s} by using an argument similar to that for Theorem~\ref{thm_sec_condition_space_general}. First, we prove the ``\,only if\,'' part by contradiction. Assume the contrary that there exists a wiretap set $W \in \mW_r$ that does not satisfy the condition \eqref{def_sec_condition_space_s}, i.e.,
  \begin{align*}
    \big\langle G_W  \big\rangle \cap \big\langle \Gamma \big\rangle \neq \big\{\vec{0}\big\}.
  \end{align*}
  With this, there exist two non-zero column vectors $\vec{\alpha} \in \Fq^{|W|}$ and $\vec{\beta} \in \Fq^{s\ell}$ such that
  \begin{align}
    G_W \cdot \vec{\alpha}= \Gamma \cdot \vec{\beta} \neq \vec{0}.\label{eq_HWw=Gav_s}
  \end{align}
  We now consider
  \begin{align}
        &I\big(\vM_S ; \vY_W\big) = I\big(\vM_S ; (\vM_S~\vK_S)  \cdot G_W\big)\nonumber\\
        &\geq I\big(\vM_S;\, (\vM_S~\vK_S) \cdot G_W \cdot \vec{\alpha}\big)\nonumber\\
        &= I\big(\vM_S;\,(\vM_S~\vK_S) \cdot \Gamma \cdot \vec{\beta} \,\big) \label{eq_Gav_s}\\
        &= I\big(\vM_S;\, \vM_S \cdot \vec{\beta}\, \big)\label{eq_HMLd*=0_s}\\
        &=H\big(\vM_S \cdot \vec{\beta}\, \big) - H\big(\vM_S \cdot \vec{\beta} \,\big|\, \vM_S\big)\nonumber\\
        &= H\big(\vM_S \cdot \vec{\beta}\, \big)\nonumber\\
        & > 0, \label{eq_If_fb>0_s}
    \end{align}
  where the equalities \eqref{eq_Gav_s} and \eqref{eq_HMLd*=0_s} follows from \eqref{eq_HWw=Gav_s} and \eqref{eq_XSGA=fMS_s}, respectively; and \eqref{eq_If_fb>0_s} follows from $\vec{\beta} \neq 0$. Thus we have obtained that $I\big(\vM_S ; \vY_W\big) > 0$, which contradicts the source-security condition \eqref{eq_equivalent_source_security_condition_s}.

  For the ``\,if\,'' part, it suffices to prove that for each $W \in \mW_r$, the equality
  \begin{align}
    \Pr\big(\vM_S = \vf \,\big|\, \vY_W = \vy\big) =  \Pr\big(\vM_S = \vf\big)\label{eq_x_given_y=x_s}
  \end{align}
  holds for any row vector $\vf \in \Fq^{s\ell}$ and any row vector $\vy \in \Fq^{|W|}$ with $\Pr\big(\vY_W = \vy\big) >0$, namely that there exists a pair~$(\vm_S~\vk_S)$ of a vector of source messages $\vm_S$ and a vector of keys $\vk_S$ such that $$(\vm_S~~\vk_S) \cdot \Big[\vg_e:~e\in W\Big] = \vy.$$

We first note that
\begin{align}\label{eq_Prf=uniform_s}
  \Pr \big( \vM_S = \vf) = \frac{1}{q^{s\ell}}, \quad\forall~\vf \in \Fq^{s\ell}.
\end{align}
Now consider
\begin{align}
  & \Pr\big( \vM_S = \vf \,\big|\, \vY_W = \vy \big) \nonumber\\
  &= \frac{\Pr\big(\vM_S= \vf, \vY_W = \vy\big)}{\Pr\big( \vY_W = \vy\big)} \nonumber\\
  &=\frac{\Pr\big((\vM_S~\vK_S) \cdot \Gamma = \vf, (\vM_S~\vK_S) \cdot G_W = \vy\big)}{\Pr\big((\vM_S~\vK_S) \cdot G_W = \vy\big)}\nonumber\\
  &=\frac{\Pr\big((\vM_S~\vK_S) \cdot \big[\Gamma~G_W \big] = (\vf~\vy)\big)}{\Pr\big((\vM_S~\vK_S) \cdot G_W = \vy\big)}\nonumber\\
  &=\frac{\sum_{(\vm_S~\vk_S):(\vm_S~\vk_S) \cdot [\Gamma~G_W] = (\vf~\vy)}\Pr\big(\vM_S=\vm_S,\vK_S=\vk_S) \big)}{\sum_{(\vm_S'~\vk_S'):~(\vm_S'~\vk_S') \cdot G_W = \vy}\Pr\big(\vM_S=\vm_S', \vK_S=\vk_S')\big)}\nonumber\\
  &= \frac{\#\big\{(\vm_S~\vk_S):(\vm_S~\vk_S) \cdot [\Gamma~G_W] = (\vf~~\vy)\big\}}
                  {\# \big\{(\vm_S'~\vk_S'): (\vm_S'~\vk_S') \cdot G_W = \vy\big\}} \label{eq_sec_cond_prob_2_s}.
\end{align}
Furthermore, for the dominator of \eqref{eq_sec_cond_prob_2_s}, we have
\begin{align}
  \#\Big\{(\vm_S'~\vk_S'):~ (\vm_S'~\vk_S') \cdot G_W = \vy\Big\} = q^{(s\ell + \sum_{i=1}^{s}z_i) - \Rank(G_W)}\label{eq_RGW_s};
\end{align}
and for the numerator of \eqref{eq_sec_cond_prob_2_s}, we have
  \begin{align}
    &\#\Big\{(\vm_S~~\vk_S): ~(\vm_S~\vk_S) \cdot \big[\Gamma~G_W \big] = (\vf~~\vy)\Big\}\nonumber\\
    &= q^{(s\ell + \sum_{i=1}^{s}z_i) - \Rank([\Gamma~G_W])} \nonumber\\
    &= q^{(s\ell + \sum_{i=1}^{s}z_i) - \Rank(\Gamma) - \Rank(G_W)} \label{eq_RGammaGW_s}\\
    & =q^{(s\ell + \sum_{i=1}^{s}z_i) - s\ell - \Rank(G_W)}, \label{eq_RGammaGW_s-2}
  \end{align}
where \eqref{eq_RGammaGW_s} follows from $\Rank\big([\Gamma~G_W]\big) = \Rank(\Gamma) + \Rank(G_W)$ by \eqref{def_sec_condition_space_s}, and \eqref{eq_RGammaGW_s-2} follows from $\Rank(\Gamma) = s\ell$ by \eqref{eq_GammaCs_def_s}. Combining \eqref{eq_Prf=uniform_s}, \eqref{eq_sec_cond_prob_2_s}, \eqref{eq_RGW_s} and \eqref{eq_RGammaGW_s}, we have proved the equality \eqref{eq_x_given_y=x_s}, i.e.,
\begin{align*}
  \Pr\big(\vM_S = \vf \,\big|\, \vY_W = \vy\big) = \frac{1}{q^{s\ell}}=  \Pr\big(\vM_S = \vf\big).
\end{align*}
This proves Theorem~\ref{thm_sec_condition_space_general_for_s}.

\section{One-to-One Correspondence of Matrices in $\widehat{\mB}$ and Codes Constructed\\ by the Code Construction}\label{app_2}

We consider the target-function-security model $\langle \mN, f, r\rangle$ with security level $0 \leq r \leq C_{\min}$. Let $\mbC$ be an $\Fq$-valued admissible $(R, 1)$ linear secure network code for the model $(\mN, f)$, where $R$ is an integer with $r \leq R \leq C_{\min}$. In this appendix we will prove that the matrices $\widehat{B} \in \widehat{\mB}$ (cf.~\eqref{eq_Bhat_set} for $\widehat{\mB}$) and the $\Fq$-valued admissible $(R-r, 1)$ linear secure network codes $\hmbC$ for $\langle\mN, f, r\rangle$ that can be constructed by our code construction in Section~\ref{subsec_code_construction} are one-to-one corresponding. By \textbf{Verification of the Computability Condition} and \textbf{Verification of the Target-Function-Security Condition} in Section~\ref{subsec_code_construction}, we can readily see that each matrix $\widehat{B}$ obtained by our code construction (cf.~Section~\ref{subsec_exis_b} for obtaining $\widehat{B}$) satisfies the conditions \eqref{eq_bR_in_Grho} and \eqref{eq_bR_cap_G=0}, i.e.,
\begin{equation}\label{eq_bR_in_Grho_app}
    \big\langle \widehat{B}^{-1} \cdot \matT  \big\rangle \subseteq \big\langle H_\rho \big\rangle,
\end{equation}
and
\begin{equation}\label{eq_bR_cap_G=0_app}
    \big\langle \widehat{B}^{-1} \cdot \matT  \big\rangle \cap \big\langle  H_W \big\rangle = \{\vec{0}\}.
  \end{equation}
Hence, it suffices to prove that each matrix $\widehat{B} \in \widehat{\mathcal{B}}$ can be obtained following the construction in Section~\ref{subsec_exis_b}.

We consider an arbitrary matrix
\begin{align}\label{equ97}
\widehat{B} =\Big[ B_1~~B_2~~\cdots~~B_s  \Big]_s^{\text{diagonal}}
\end{align}
in $\widehat{\mB}$. By the invertibility of the matrices $B_1, B_2, \cdots, B_s$, we only need to prove that \eqref{vb_j-1} is satisfied, i.e.,
  \begin{align}\label{eq_vbj_equivalence_app}
\vb_j \in
 \big\langle H_\rho\big\rangle
 \setminus \left[ \left( \bigcup_{W \in \mW_r}(\mH_W + \mB_{j-1})\right) \cup \left( \bigcup_{i=1}^s \mathcal{D}^{\langle i \rangle}_{j-1} \right) \right],~~\forall~1 \leq j \leq R-r,
\end{align}
where we recall that
\begin{equation}\label{equ99}
  \vb_j = \begin{bmatrix}
  \vb^{\,(1)}_j \\
  \vb^{\,(2)}_j \\
\vdots \\
\vb^{\,(s)}_j
\end{bmatrix}~~\text{and}~~B_i^{-1} = \Big[\vb^{\,(i)}_1 ~~ \vb^{\,(i)}_2 ~~ \cdots ~~\vb^{\,(i)}_R \Big],\quad \forall~1\leq i \leq s.
\end{equation}
Further, from \eqref{equ97}, \eqref{equ99} and \eqref{eq_T_in_construct} for $T$,  we obtain
\begin{equation*}
    \widehat{B}^{-1} \cdot \matT =  \Big[\vb_1~~\vb_2~~\cdots~~\vb_{R-r}\Big],
\end{equation*}
which, together with \eqref{eq_bR_in_Grho_app}, implies that $\vb_j \in \langle H_\rho \rangle$ for each $1\leq j \leq R-r$. It follows from the invertibility of the matrices $B_1, B_2, \cdots, B_s$ that $\vb^{\,(i)}_j$ is linearly independent of $\vb^{\,(i)}_1, ~\vb^{\,(i)}_2, \cdots, \vb^{\,(i)}_{j-1}$ for each $1\leq i \leq s$ and $1\leq j \leq R-r$. Thus this implies that $\vb_j \notin \bigcup\limits_{i=1}^s \mathcal{D}^{\langle i \rangle}_{j-1}$ and $\vb_j \notin \mB_{j-1}$ for each $1\leq j \leq R-r$. We now rewrite \eqref{eq_bR_cap_G=0_app} as
\begin{equation*}
  \{\vec{0}\} = \big\langle \widehat{B}^{-1} \cdot \matT  \big\rangle \cap   \big\langle  H_W \big\rangle = \big\langle\vb_1,~\vb_2,\cdots,\vb_{R-r} \big\rangle \cap \big\langle H_W \big\rangle,~~\forall~W \in \mW_r,
\end{equation*}
which implies that for each $1 \leq j \leq R-r$,
\begin{equation*}
  \mB_j \cap \mH_W = \big\langle \vb_1, ~\vb_2,\cdots, \vb_j \big\rangle \cap\langle H_W\rangle = \{\vec{0}\}~~\text{and}~~\vb_j \notin\langle H_W\rangle,~~\forall~W \in \mW_r.
\end{equation*}
Thus, for $1 \leq j\leq R-r$, we have
\begin{equation*}
  \vb_j \notin \mH_W + \mB_{j-1},~~\forall~W \in \mW_r.
\end{equation*}
Combining the above, we see that
\begin{equation*}
  \vb_j \notin \left( \bigcup_{W \in \mW_r}(\mH_W + \mB_{j-1}) \right) \cup \left( \bigcup_{i=1}^s \mathcal{D}^{\langle i \rangle}_{j-1}\right),\quad \forall~1\leq j \leq R-r.
\end{equation*}
Together with $\vb_j \in \langle H_\rho \rangle$ for all $1\leq j \leq R-r$, we thus have proved \eqref{eq_vbj_equivalence_app}. Then, we have completed the proof that the matrices in $\widehat{\mB}$ and the codes constructed by our code construction are one-to-one corresponding.


\end{document}